%% file: ms.tex
\documentclass[a4paper,fleqn,usenatbib]{mnras}

\usepackage{newtxtext,newtxmath}

\usepackage[T1]{fontenc}
\usepackage{ae,aecompl}

\usepackage{graphicx}	
\usepackage{amsmath}	
\usepackage{amssymb}	
\usepackage{wasysym}
\usepackage{verbatim}  
\usepackage{vector}  

\title[Disruption of Hierarchical Triplets]{A Simple Random-Walk Model Explains the Disruption Process of Hierarchical, Eccentric 3-Body Systems}

\author[J. Mushkin and B. Katz]{
Jonathan Mushkin,$^{1}$\thanks{E-mail: jonathan.mushkin@weizmann.ac.il (JM)}
Boaz Katz$^{1}$
\\
$^{1}$Department of Particle Physics and Astrophysics, Weizmann Institute of Science, Rehovot 76100, Israel
}

\date{Accepted XXX. Received YYY; in original form ZZZ}

\pubyear{2019}

\begin{document}
\label{firstpage}
\pagerange{\pageref{firstpage}--\pageref{lastpage}}
\maketitle

\begin{abstract}
We study the disruption process of hierarchical 3-body systems with bodies of comparable mass. Such systems have long survival times that vary by orders of magnitude depending on the initial conditions. By comparing with 3-body numerical integrations, we show that the evolution and disruption of such systems can be statistically described as a simple random-walk process in the outer-orbit's energy, where the energy-exchange per pericenter passage (step-size) is calculated from the initial conditions. In our derivation of the step-size, we use previous analytic results for parabolic encounters, and average over the (Kozai-Lidov) oscillations in orbital parameters, which are faster then the energy diffusion timescale. While similar random-walk models were studied before, this work differs in two manners: (a) this is the first time that the Kozai-Lidov averaged step-size is derived from first principles and demonstrated to reproduce the statistical evolution of numerical ensembles without fitting parameters, and (b) it provides a characteristic life-time, instead of answering the binary question (stable/unstable), set by case-specific criteria.
\end{abstract}




\input{Sections/Introduction} 
\input{Sections/NumericalSimulations}
\input{Sections/RandomWalkModel}

\input{Sections/Conclusions}

\section*{Acknowledgements}
The authors would like to thank Subo Dong and Mario Livio for helpful discussions.



\bibliographystyle{mnras}
\bibliography{Bibliography}



\appendix
\input{Sections/Appendix1}
\input{Sections/Appendix2}
\input{Sections/Appendix3}

\bsp	
\label{lastpage}
\end{document}

%% file: Sections/Introduction.tex
\section{Introduction}
\label{sec:introduction}
Hierarchical three body systems are ubiquitous among astrophysical systems.
The subject of this work is the gravitational three-body problem. It has occupied the minds of scientists for hundreds of years \citep[see e.g.][for a recent review]{valtonen2006three}. Hierarchical triple systems of comparable masses, where two of the bodies are relatively close, and the third body is relatively distant form them, are particularly interesting due to their long-term stability. Most observed 3-body stellar systems are hierarchical for the simple reason that other systems have been disrupted quickly after formation. In such systems, angular momentum is exchanged between the inner binary (comprising the two inner bodies) and the outer binary (comprising of the inner binary and the outer body) more efficiently than energy, leading to intermediate-time-scale oscillations in the orbital elements \citep[Kozai-Lidov oscillations,][]{lidov1962evolution, kozai1962secular}. While the energy exchange is slower than the angular momentum exchange, it can accumulate and affect the hierarchy of the system or lead to a disruption over long-time scales, and is thus important for studying the distribution of existing stellar systems.

A common definition of a system's stability is by whether or not one body is ejected, or there is exchange between members of the inner and outer binaries, before $N$ outer orbits are completed. Many works have dealt with phrasing a stability criterion, such that system which obtain it are (almost) certainly stable \cite[for detailed reviews, including other definitions of stability, see][]{valtonen2006three, georgakarakos_stability_2008}.  Some criteria \citep[e.g.][]{mardling2001tidal, myllari2018stability} have a practical motivation: in many-body ($\gtrsim 10^4$) simulations, each stable triple systems can be treated with computationally cheaper tools. Often, the criteria are purely empirical \citep[e.g.][]{harrington_stability_1972, eggleton1995empirical}. Several recent works \citep{valtonen2006three, valtonen2008stability, myllari2018stability} used a semi-analytical approach, in which the approximated analytic energy exchange formula derived by \citet[][also equation  \eqref{RHdeltaEltaEsimple} below]{roy2003energy} is averaged to produce a step-size for a random walk model in the outer binary's energy, and empirical fitting is used to determine the exact criterion. All empirical criteria have the disadvantage of convolving arbitrary time-scales for stability with fitted functions and factors. If the criterion was derived to consider stability for $10^{4}$ outer orbits, it is not trivial how should one adjust it for $10^{6}$ outer orbits, for example.

In this paper, we study the disruption process as a diffusion, and find the suitable way to characterize it. This is an extension to the previous works mentioned in two aspects: we discuss stability in a continuous sense, not through an arbitrary timescale; and derive our results without any empirical fitting. The paper is structured as follows. In \S \ref{sec:NumericalSimulations} we present the trends and regularity in the "lifetime", in terms of outer-orbit revolutions and years, of equal mass triple systems with high outer-orbit eccentricity. In \S \ref{sec:RWmodel} the Random Walk model is derived and compared with full 3-body integrations, in terms of energy exchanges and "lifetime", the latter being the main results of this work (Figures \ref{fig:LifetimeToPericenter} and \ref{fig:LifetimeToIncination}). In \S\ref{sec:summary} the work is concluded. 

Throughout this paper we use the following notations. The hierarchical triple systems consists of an \textit{inner orbit} consisting of masses $m_1,m_2$ and an \textit{outer orbit} consisting of the distant third mass $m_3$ and the center of mass of the inner orbit. The main parameters of the each orbit that are used are the semi-major axis $a$, energy $E$, period $P$, eccentricity $\bvec{e}$, pericenter distance $r_{\rm p}=a(1-e)$, angular momentum $\bvec{J}$, total mass $M$, reduced mass $\mu$. The phases of the orbits are described using the mean anomaly ${\rm M}$. Orientation is set by inclination $\iota$ (angle between $\bvec{J}$ and the $z$-axis), longitude of ascending node $\Omega$ and argument of periapsis $\omega$. Subscripts $"\textrm{i}"$ and $"\textrm{o}"$ to denote quantities that are related to the inner and outer orbits respectively. Adopting from \citet{roy2003energy}, and unlike common practice in Kozai-Lidov framework, $\iota$, $\Omega$ and $\omega$ without subscripts are of the outer orbit, in a coordinate system set by the inner orbit, with $\hat{\bvec{x}}\parallel \bvec{e}_{\rm i}$ and $\hat{\bvec{z}}\parallel\bvec{J}_{\rm i}$.

%% file: Sections/NumericalSimulations.tex
\section{Numerical Simulations Results for Equal Mass Hierarchical Triplets with Eccentric Outer Orbits}
\label{sec:NumericalSimulations}
The simulations presented in this work were performed using \mbox{\textsc{HopOn}}, a dedicated PYTHON 3.6 package written by the authors that is described in Appendix \ref{subsec:AdaptiveLeapFrog}. 

We have performed 4,000 simulations of hierarchical triple systems with the following parameters (summarized in Table \ref{tab:ic}) : the masses of the three bodies are equal, the outer-orbit's eccentricity is set to $e_{\textrm{o}}=0.9$, the inner-orbit's eccentricity is set to $e_{\textrm{i}}=0.5$, the hierarchy is randomly (uniformly) chosen in the range $r_{\textrm{p.o}}/a_{\textrm{i}}=2.0\textrm{-}4.5$, random isotropic relative orientation and random inner mean anomaly. The simulations all begin with the outer orbit at apocenter ($\mathrm{M}_{\mathrm{o}}=\pi$). Each simulation is terminated when the system is disrupted (one body moves away from the other two to sufficiently large distance with positive energy) or when reaching the limiting run time of $10^{9}$ time steps. Criteria termination are given in appendix \ref{subsec:AdaptiveLeapFrog}. 

In figure \ref{fig:LifetimeToPericenter} we present the lifetime up to disruption of the simulations, measured in completed outer orbits $N_{\textrm{o.o}}$ (upper panel) and in physical time passed before disruption $T$ (on the bottom panel, for a particular choice an inner period of 300 years). Red dots represent systems disrupted within our computational run-time constraints. Black triangles show the run time of simulations that were terminated before disruption, providing a lower limit to the lifetime of these systems. Blue dots (gray solid lines) are the results of detailed (simplified) random-walk analytic models discussed in section \ref{sec:RWmodel}. 

As can be seen in figure \ref{fig:LifetimeToPericenter}, the lifetime of the systems grows rapidly (faster than exponentially) for increasing $r_{\textrm{p.o}}/a_{\textrm{i}}$ with striking regularity. The rising trend is expected \citep[e.g.][]{harrington_stability_1972}, due to the smaller energy exchange between the outer mass and the binary for larger separations. The regularity suggests a simple origin which we show below is largely explained by a simple random-walk behaviour. 

\begin{figure}
	\centering
	\includegraphics[width=0.4\textwidth]{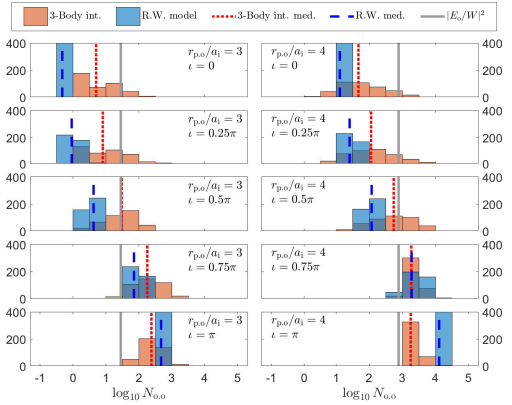}
	\caption{Disruption time of equal mass hierarchical eccentric triplets, with specific initial outer pericenter distance $r_{\textrm{p.o}}$ and inclinations $\iota$ (see Table \ref{tab:ic}). Each panel shows the distribution of $N_{\rm o.o}$ for 400 systems initial  $e_{\textrm{o}}=0.9$, $e_{\textrm{i}}=0.5$, random angles $\Omega$ and $\omega$, random ${\rm M_i}$, $r_{\textrm{p.o}}/a_{\textrm{i}}=3$ (left panels)  or 4 (right panels), and 5 equally spaced $\iota$ values (top to bottom): $0$, $\pi/4$, $\pi/2$, $3\pi/4$ and $\pi$. Red bars: 3-body integrations. Blue bars: Random Walk model, for same initial conditions. Red dotted line: median value of 3-body integrations. Blue dashed line: median value of Random Walk model results. Solid gray line: rough analytic random-walk estimate, ignoring the dependence on orientation and inner eccentricity.}
	\label{fig:LifetimeToIncination}
\end{figure}

A significant part of the observed scatter is due to the dependence of the energy exchange on the relative inclination between the inner and outer orbits \citep[e.g.][]{myllari2018stability}, as can be seen in the red bars in the panels of figure \ref{fig:LifetimeToIncination}. For this figure, we performed 400 simulations with random orientation and relative phases, for each of 2 values of $r_{\rm p.o}$ and 5 equally spaced inclination values, as described in Table \ref{tab:ic}. Those results are not separated into resolved and unresolved. The rest of the scatter is attributed to the other randomly selected variables.

An example simulation can be seen in figure \ref{fig:3BodyExample}. On the top panel, we see that the outer and inner orbits exchange energy, until the system disrupts at $E_{\textrm{o}}\geq0$. The lower panel focuses on the early part of the integration of the system. As can be seen, the evolution of $e_{i}$ and the relative inclination undergo periodic Kozai-Lidov oscillations, while the trajectory of $E_{\textrm{o}}$ is a sum of what appears to be random contributions. In fact, as the amplitude of the energy exchanges depend on oscillating orbital parameters, the Kozai-Lidov oscillations affect the energy exchange. A hint for this dependence is seen in the figure with larger typical exchanges at phases with low inclination and high-eccentricity. Given that the evolution of Kozai-Lidov oscillations is known analytically, their effect on the energy exchange is calculated and incorporated into the random-walk model in sections \S\ref{sec:RWmodel} (see in particular figure \ref{fig:deltaEMeanAndVariance}).

\begin{figure}
	\centering
	\begin{minipage}{0.40\textwidth}
		\includegraphics[width=\textwidth, trim={0 0.2cm 0 1cm}, clip]{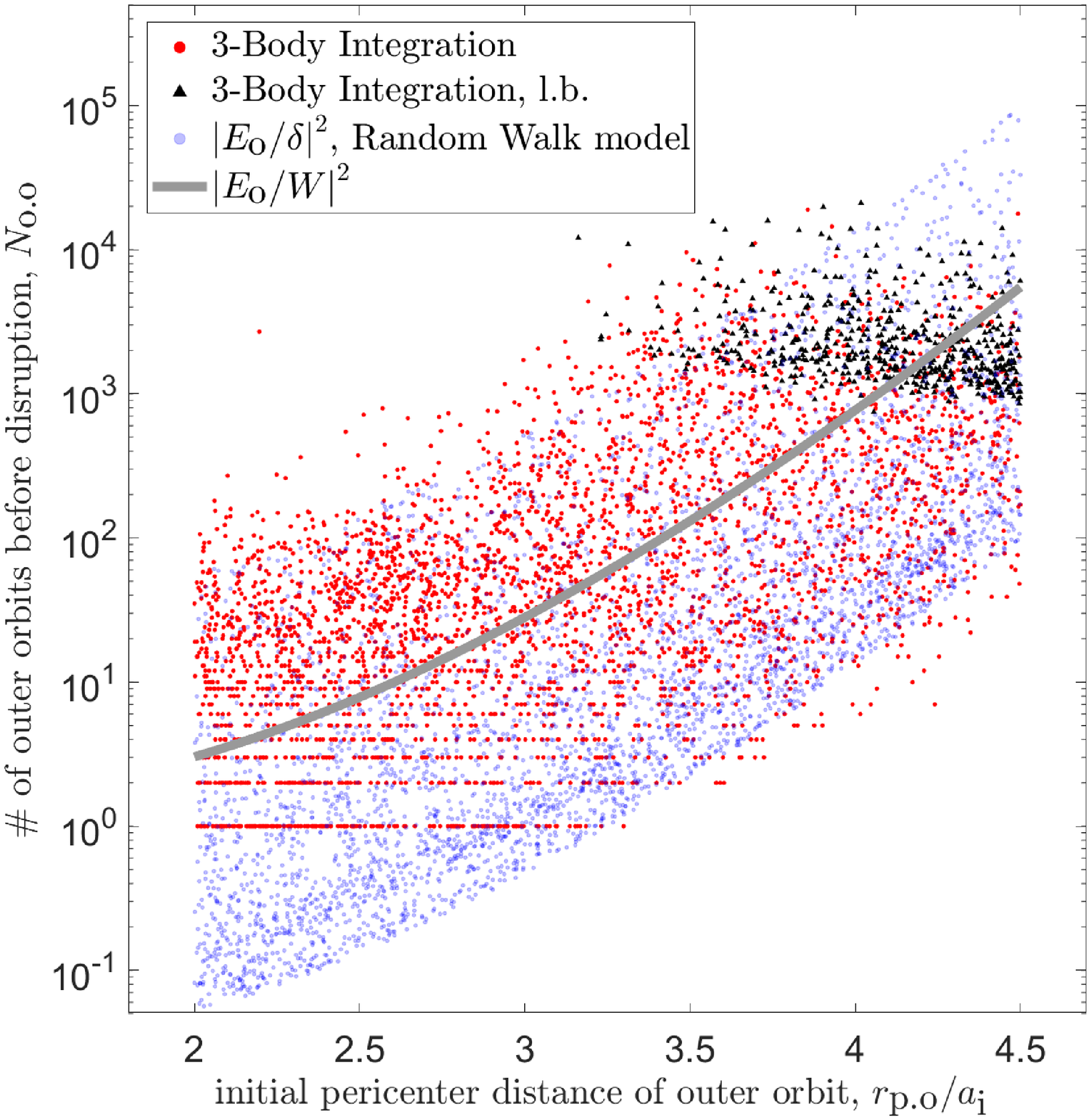}
	\end{minipage}
	\hfil
	\begin{minipage}{0.40\textwidth}
		\includegraphics[width=\textwidth, trim={0 0.2cm 0 1cm},clip]{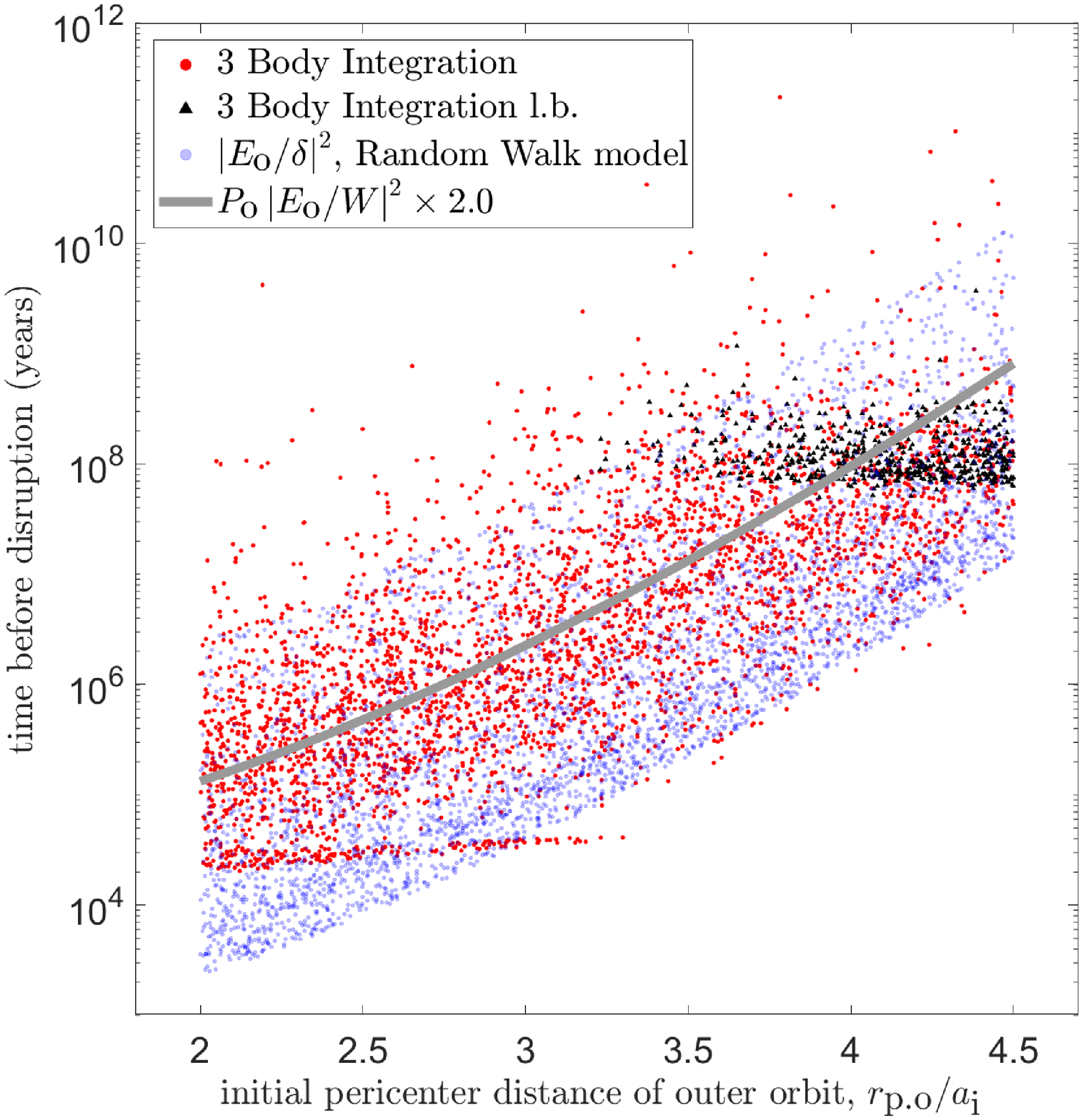}
	\end{minipage}
	\caption{
		Disruption time of equal mass hierarchical eccentric triplets. Data points represent 4,000 systems with initial $e_{\rm o}=0.9$, $e_{\textrm{i}}=0.5$, $r_{\rm p.o}/a_{\textrm{i}}=2\textrm{-}4.5$, and random isotropic orientations (see Table \ref{tab:ic}).  Red dots: numerical 3-body integrations. Black triangles: lower bounds, from systems undisrupted at end of simulation. Blue dots: expected outer orbits  for each of the same initial conditions, based on a random walk model (equation \eqref{eq:NooRW}) with step sizes using the approximate energy exchange averaged over Kozai-Lidov cycles and orbital phases (equation \ref{eq:RWdelta}). Grey line: Rough analytic random-walk estimate ignoring the dependence on orientation and inner eccentricity (Equation \ref{eq:RoughNooEst}). The line captures the trend of the red and blue dots, but not their scatter. \textbf{Upper panel}: number of outer orbit revolutions performed until the systems disrupt, $N_{\textrm{o.o}}$. \textbf{Lower panel}: time passed until the systems disrupt, for inner binaries with $m_1=m_2=1 M_{\astrosun}$ and periods of $300$ years. Time estimation is described in Section \ref{subsec:T}.
	}	
	\label{fig:LifetimeToPericenter}
\end{figure}

%% file: Sections/RandomWalkModel.tex
\section{The Random-Walk Model for 3-Body System Evolution}
\label{sec:RWmodel}
\begin{figure}
	\centering
	\begin{minipage}{0.49\textwidth}
		\centering
		\includegraphics[width=\textwidth,  trim={0 0.0cm 0 0cm}, clip]{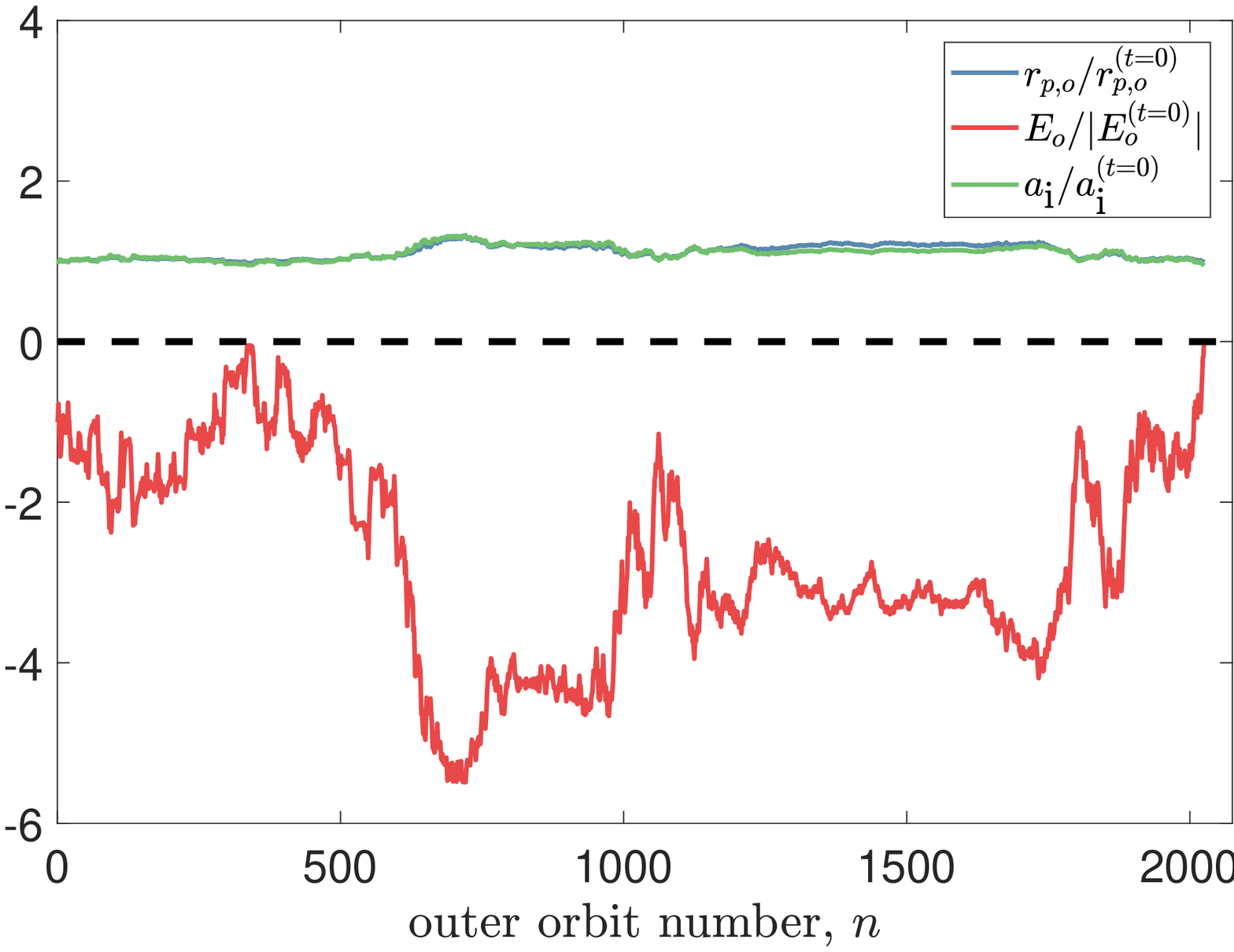}
		\label{fig:3BodyExampleLong}
	\end{minipage}
	\hfill
	\begin{minipage}{0.49\textwidth}
		\centering
		\includegraphics[width=\textwidth,  trim={0 0.0cm 0 0cm}, clip]{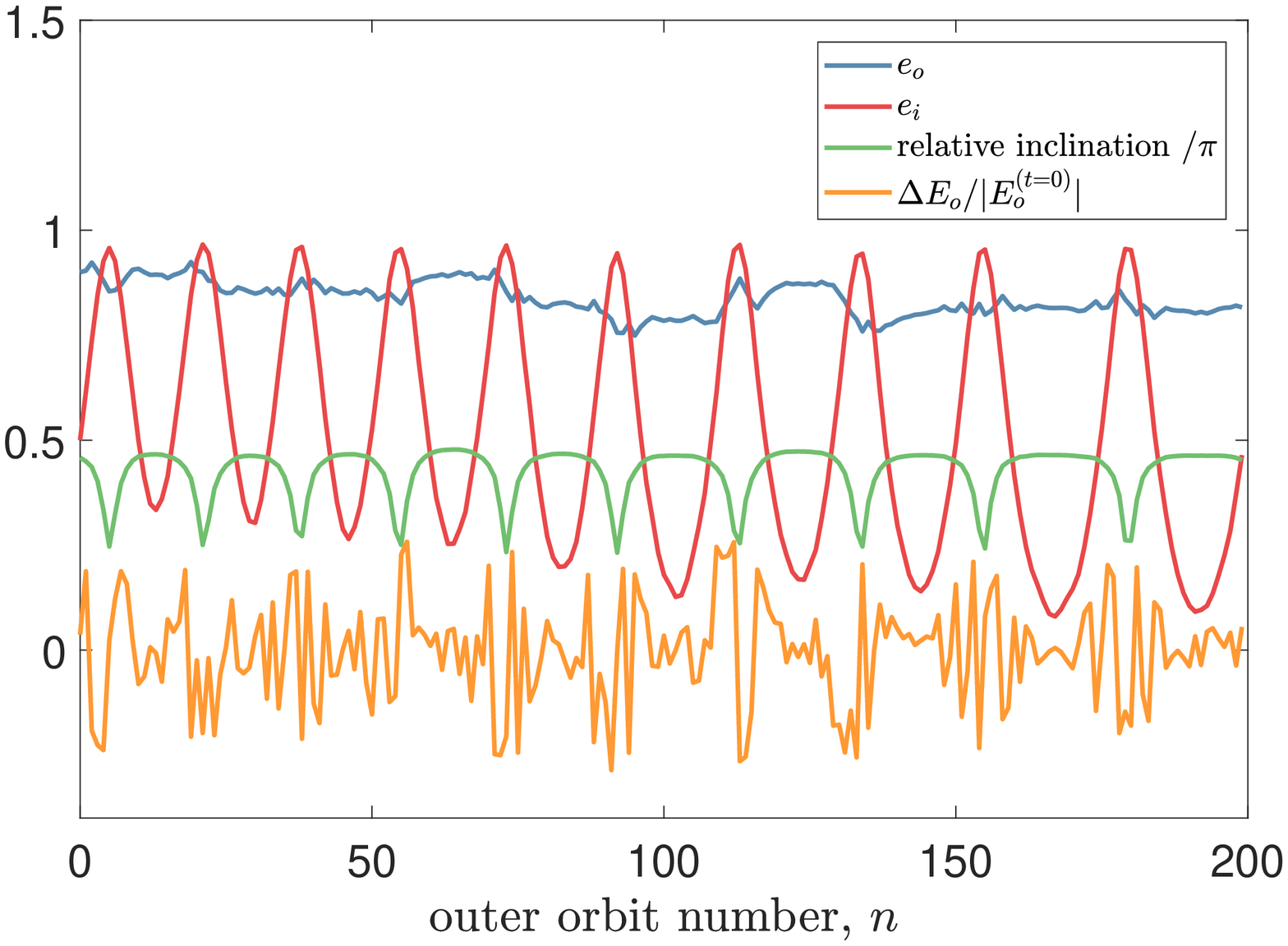}
		\label{fig:3BodyExampleShort}
	\end{minipage}
	\hfill
	\caption{An example of the long and short term trends in the evolution of an hierarchical ($a_{\textrm{o}}/a_{\textrm{i}}\approx43.3$), eccentric ($e_{\textrm{o}}=0.9$) equal mass triple system (from simulations specified in table \ref{tab:ic}), produced with \textsc{HopOn} 3-body integrator. Data points are taken at each outer apocenter passages.
		\textbf{Top panel}: entire lifetime. Blue: outer orbit's pericenter distance. Red: outer orbit's energy. Green: inner orbit's semi-major axis. Curves are normalized by their initial values.
		The system is disrupted when $E_{\textrm{o}}=0$ (black dashed line).
		\textbf{Bottom panel}: first 200 outer orbits of same system. Blue: eccentricity of outer orbit. Red: eccentricity of inner orbit. Green: relative inclination between the two orbits. Yellow: energy exchange per orbit, normalized by initial $E_{\textrm{o}}$ value.}
	\label{fig:3BodyExample}
\end{figure}

We propose a simple model to describe the dynamics that an eccentric, mildly hierarchical triplet of comparable masses will undergo during its disruption process. There are three time-scales involved in this model: Within a single outer orbit pericenter passage, a small amount of energy is exchanged. Within $\sim 10$ outer orbits, the Kozai-Lidov oscillations forces an exchange of angular momentum between the inner and outer orbits, changing the orbital parameters periodically \footnote{Strictly speaking, the Kozai-Lidov oscillations are not periodic, due to the percession of the outer orbit within its own plane. However, the approximate energy exchange variance averaged over inner phase, is invariant to such percession, due to the dependence on this orientation through Equation \eqref{eq:phisimple}.}. Only after many outer orbits the energy changes accumulate considerably. All three time scales can be seen in the example in Figure \ref{fig:3BodyExample}. Our model takes into account the intermediate time-scales physics, and determines a single, constant typical energy exchange size, $\delta$, that is used for the Random-Walk on the outer orbit's energy. 
In section \S\ref{subsec:EnergyExchange} the approximated expression for the energy exchange in a single parabolic passage \citet{roy2003energy} are provided. In section \S\ref{subsec:RWstepsize} these expressions are averaged over a  Kozai-Lidov period numerically, providing the step-size used in the Random-Walk model. In section \S\ref{subsec:T} the averaged step-sizes are used to derive expressions for the lifetimes of hierarchical triple systems, using different levels of simplification. The results of the model are compared to numerical experiments in section \S\ref{subsec:comparison}.

\subsection{Energy Exchange in a single pericenter passage}
\label{subsec:EnergyExchange}
The simple random walk model focuses on eccentric outer orbits, and the energy exchange is estimated in the parabolic limit ($e_{\mathrm{o}}=1$).  We use the approximate analytic expressions derived by \citet{roy2003energy}. These expressions involve several approximations and agree with numerical experiments within $\apprle 25\%$. In particular the tidal force's work is calculated along unperturbed (Keplerian) orbits, neglecting changes in the trajectories during the interaction. It is useful to express the resulting energy exchange in the following way (see \S\ref{subsec:EnergyExchangeApp})
\begin{align}
\label{RHdeltaEltaEsimple}
\Delta E_{\textrm{o}} &= W(a_{\textrm{i}},r_{\textrm{p.o}},m_{1},m_{2},m_{3}) F(\phi,\Omega,\iota,e_{\textrm{i}}),\\
F &= \sqrt{2}A_{1} \sin\phi + 2A_{2}\sin\phi\cos2\Omega + 2 A_{3}\cos\phi\sin2\Omega\\
\label{eq:Wsimple}
W &= -E_{\textrm{i}} \frac{m_{3}}{M_{\textrm{i}}}\left(\frac{M_{\textrm{i}}}{M_{\textrm{o}}}\right)^{5/4} \left(\frac{r_{\textrm{p.o}}}{a_{\textrm{i}}}\right)^{3/4} e^{-2K/3},\\
K &= \left(\frac{r_{\textrm{p.o}}}{a_{\textrm{i}}}\right)^{3/2}\left(\frac{2M_{\textrm{i}}}{M_{\textrm{o}}}\right)^{1/2},\\
\label{eq:phisimple}
\phi &= 2\omega - \textrm{M}_{\textrm{i}}^{*},
\end{align}
where $\textrm{M}_{\textrm{i}}^{*}$ is the mean anomaly of the inner orbit during the next outer orbit periapsis (calculated for the unperturbed orbits), and $A_{1-3}$ are functions of $\iota$ and $e_{\mathrm{i}}$ and are given in \eqref{eq:A1A2A3}-\eqref{eq:f1f2f4}. Note that Equation \eqref{RHdeltaEltaEsimple} is derived for non-circular inner orbits, and separate expressions were derived by \citet{roy2003energy} for cases with circular inner orbits. For the systems considered here, near-circular orbits occur rarely and for short periods, and we ignore this caviat and use Equation \eqref{RHdeltaEltaEsimple} in all cases.

\subsection{Averaged energy exchange, used in the random walk model $\delta$}
\label{subsec:RWstepsize}
Random Walk models are based on the assumption that the inner mean anomaly changes randomly between outer pericenter passages and that the average energy exchange is zero \citep[e.g.][ and demonstrated at the end of this section]{myllari2018stability}. The random walk step-size is set by the variance of the energy exchange. Due to the secular evolution, the variance oscillates in time (see example in Figure \ref{fig:3BodyExample}, bottom panel). To a leading approximation, the secular evolution can be calculated by expanding the perturbing Hamiltonian to quadrupole order and averaging the equations of motion over the inner and outer periods \citep{lidov1962evolution, kozai1962secular}. Within this Double Average (DA) approximation, $\iota$, $\Omega$ and $e_{\textrm{i}}$ change periodically \footnote{Note that $\omega$ does not change periodically, but does not effect the variance of the energy exchange.}. The effective random walk step-size is therefore estimated by averaging the variance over these oscillations.

Using Equation \eqref{RHdeltaEltaEsimple}, the variance of energy exchange  (for random $\rm M^*_{\rm i}$ or, equivalently, $\phi$) is given by 
\begin{equation}
\label{eq:delta2PhaseAveraged}
\langle\Delta E_{\rm o}^2\rangle_{\rm M^*_{\rm i}} = W^2\bigg( A_1^2 + 2 A_2^2 \cos^2 2\Omega + 2 A_3^2 \sin^2 2\Omega +2\sqrt{2}A_1 A_2 \cos 2\Omega \bigg).
\end{equation}
The step-size of the random walk model is evaluated by averaging Equation \eqref{eq:delta2PhaseAveraged} over a Kozai-Lidov period $P_{\rm KL}$,
\begin{equation}
\label{eq:RWdelta}
\delta ^2 =  \int_{0}^{P_{\rm KL}} \frac{dt}{P_{\rm KL}}\langle\Delta E_{\rm o}^2\rangle_{{\rm M^*_{\rm i}}}.
\end{equation}
The averaging in \eqref{eq:RWdelta} is performed numerically by evolving the orbital parameters using the double-averaged Equations \eqref{eq:KLDA} - \eqref{eq:KLdJodt}.

\begin{figure}
	\centering
	\begin{minipage}{0.4\textwidth}
		\centering
		\includegraphics[width=\textwidth]{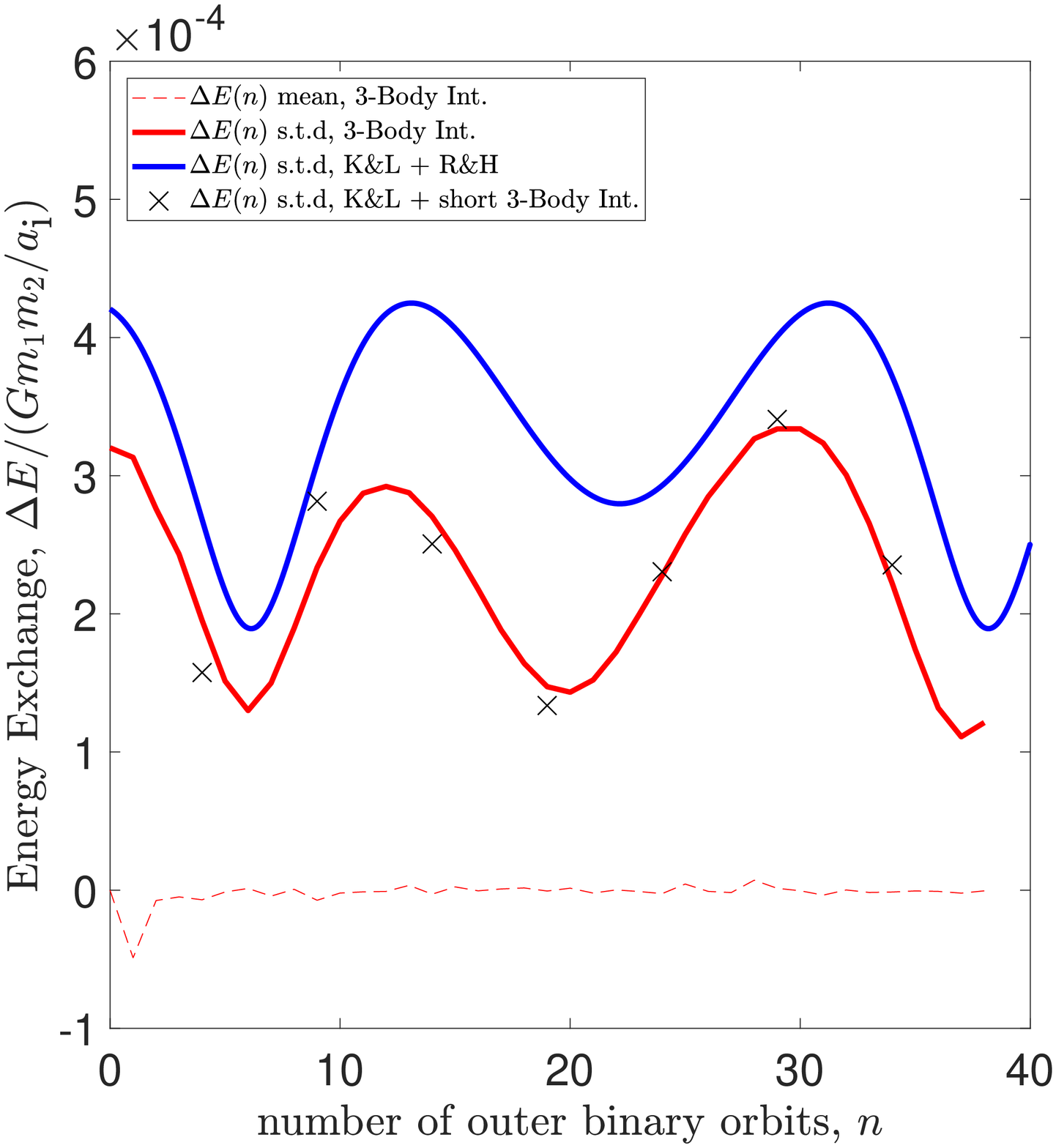}
		\label{fig:deltaEMeanAndVariance}
	\end{minipage}\hfill
	\begin{minipage}{0.4\textwidth}
		\centering
		\includegraphics[width=\textwidth]{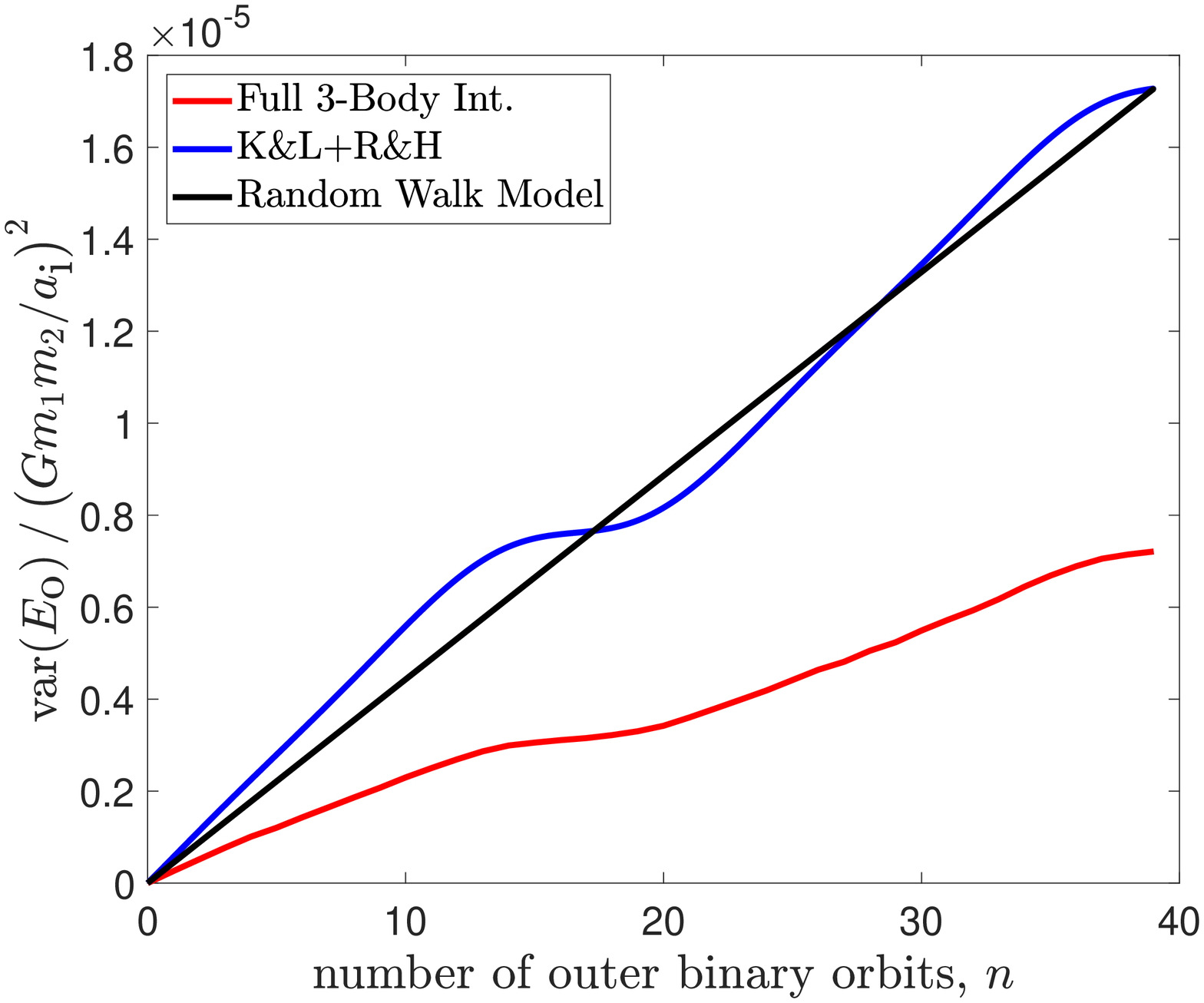}
		\label{fig:EoVarianceGrowth}
	\end{minipage}\hfill
	
	\caption{The statistical evolution of outer orbit energy ($E_{\textrm{o}}$) and energy exchange per orbit ($\Delta E_{\textrm{o}}$) for an ensemble of 10,000 identical systems of equal masses, eccentric outer orbit ($e_{\textrm{o}}=0.9$), high hierarchy ($a_{\textrm{o}}/a_{\textrm{i}}=40$), specific orientation, and varying phases (see table \ref{tab:ic}). Red curves for numerical 3-body integration, blue for Roy-Haddow exchanges calculated after Kozai-Lidov evolutions.
		\textbf{Top panel}: The mean (dashed lines) and s.t.d. (solid curve) of the exchanges $\Delta E_{\textrm{o}}$ between two sequential orbits, calculated w.r.t. $\textrm{M}_{\textrm{i}}$. Black X's: r.m.s of 1,000 short 3-body simulations, with initial conditions set by the Kozai-Lidov evolution at this point, and different inner mean anomalies. \textbf{Bottom panel}: The variance of $E_{\textrm{o}}$ as a function of number of completed outer orbits $n$. Black: $n\delta^2$ of the proposed Random Walk model, using Equation \eqref{eq:RWdelta}.}
	\label{fig:DeltaEoVarianceDemonstration}
\end{figure}

A demonstration of the validity of the averaging approach is provided Figure \ref{fig:DeltaEoVarianceDemonstration}, based on 10,000 numerical simulations of hierarchical triple systems with the following parameters (summarized in table \ref{tab:ic}) : the masses of the three bodies are equal, hierarchy set to $r_{\rm p.o}/a_{\rm i}=4$, the outer-orbit's eccentricity is set to $e_{\textrm{o}}=0.9$, the inner-orbit's eccentricity is set to $e_{\textrm{i}}=0.5$,  relative outer orbit orientation set to $\iota=3\pi/4$, $\omega= 3\pi/2$, $\Omega=\pi/5$. The simulations all begin with the outer orbit at apocenter, (${\rm M_o}=\pi$). The systems differ only by their inner mean anomaly ${\rm M_i}$, chosen randomly. The simulations were carried on for 40 outer orbits. In the upper panel of Figure \ref{fig:DeltaEoVarianceDemonstration}, the root-mean-square and mean energy exchange (solid and dashed red lines, respectively), measured between outer apocenter passages. Using the same initial conditions, we evolved a single system according to the DA Kozai-Lidov prescription, \eqref{eq:KLdeidt}-\eqref{eq:KLdJodt}, and evaluated the phase-averaged energy exchange, Equation \eqref{eq:delta2PhaseAveraged} (solid blue line). 

There is striking agreement between the averaging scheme and the direct 3-body integration (shapes of blue and red curves in Figure \ref{fig:DeltaEoVarianceDemonstration}, upper panel), up to scaling. We attribute this difference to the approximated expression used to for  $\Delta E_{\textrm{o}}$, Equation \eqref{RHdeltaEltaEsimple}. This claim is tested by using the orbital parameters set by the Kozai-Lidov evolution to perform direct 3-body integrations, and calculate the energy exchange after a single outer orbit. For each of 13 points along the evolution, 1,000 short integrations are performed, with initial inner mean anomalies evenly spaced between 0 and $2\pi$. The simulations were performed in MATLAB, using the same integrator as \textsc{HopOn} (see Appendix \ref{sec:hopon}). The root-mean-squares of the energy exchanges of each ensemble is marked with black X's in the upper panel of Figure \ref{fig:DeltaEoVarianceDemonstration}, and they agree with the curve of full 3-body integrations. We hence conclude that the separation of the evolution into independent Kozai-Lidov oscillations and random energy exchange is valid. 

The growth of the variance through time is shown in the button panel of Figure \ref{fig:DeltaEoVarianceDemonstration}. The fact that the variance ratio between the 3-body integrations and the Kozai-Lidov modulated exchanges of both $E_{\rm o}$ and $\Delta E$  is roughly the same ($\sim 0.5$) suggest there is no dominant correlations structure between consecutive exchanges, and that a random walk model is sufficient.

\subsection{Disruption Timescale Estimation}
\label{subsec:T}
The hierarchical 3-body system is disrupted once the outer energy $E_{\rm o}>0$. On the other extreme, system may become chaotic if $E_{\rm o}$ is sufficiently negative such that $a_{\rm o }\sim a_{\rm i}$. In such a case, the system usually disrupts quickly. 
The median number of steps $N_{\rm o.o}$ (outer orbits) $E_{\textrm{o}}$ performs before disruption can be therefore estimated by a random walk calculation with termination at two boundaries corresponding to $E_{\textrm{o}}\geq0$ and to $a_{\textrm{o}}\sim a_{\textrm{i}}$. $N_{\rm o.o}$ can be expressed as
{\begin{equation}
\label{eq:NooRW}
N_{\textrm{o.o}}= \alpha \left(\frac{E_{\textrm{o}}}{\delta}\right)^2,
\end{equation}
where $\alpha$ is an order-unity factor that is related to the location of the boundaries and the step size. For random walks with the wide range of boundaries and step sizes that correspond to the parameters of the simulations presented in \S\ref{sec:NumericalSimulations}, the value of $\alpha$ is in the range 0.5 - 2. Hereafter we adopt the approximation
\begin{equation}
\label{eq:alpha}
\alpha=1.
\end{equation}}
An explicit analytic approximation can be obtained, by neglecting the dependence of the energy exchange  on the orientation and inner eccentricity in Equation \eqref{RHdeltaEltaEsimple}, setting $\delta=|W|$, resulting in

\begin{equation}
\label{eq:RoughNooEst}
\begin{aligned}
N_{\textrm{o.o}} &= \left(1-e_{\textrm{o}}\right)^{2}
\left(\frac{ r_{\textrm{p.o}}}{a_{\textrm{i}}} \right) ^{-7/2}
\left(\frac{M_{\textrm{i}}}{\mu_{\textrm{i}}} \right)^{2}
\left( \frac{M_{\textrm{o}}}{M_{\textrm{i}}}\right)^{5/2} \\
&\times\exp\left(\frac{4\sqrt{2}}{3}\sqrt{\frac{M_{\textrm{i}} }{M_{\textrm{o}}} }\left(\frac{r_{\textrm{p.o}}}{a_{\textrm{i}}}\right)^{3/2}\right).
\end{aligned}
\end{equation}
While a rough approximation, Equation \eqref{eq:RoughNooEst} captures the dependence of $N_{\textrm{o.o}}$ on the masses and on $r_{\textrm{p.o}}/a_{\textrm{i}}$. 

The median lifetime of a given system experiencing random-walk in $E_{\rm o}$ can be expressed as 
{\begin{equation}
\label{eq:TRW}
T =  \beta N_{\textrm{o.o}}P_{\textrm{o}}^{(t=0)} 
\end{equation}
where $\beta$ is an order unity number, which is larger than 1 due to the fact that most of the orbits have  $E_{\mathrm{o}}$ closer  to zero and corresponding larger periods compared to the initial values. Within a random walk realization for $E_{\rm o}$, the lifetime can be easily calculated, given that $P_{\rm o}\propto \left|E_{\rm o}\right|^{-3/2}$, allowing $\beta$ to be calculated for any given random-walk boundaries and step-size. For the wide range of initial conditions presented in this work, $\beta$ is found to be in the range $2.0$-$3.0$. Henceforth, we adopt the approximation\footnote{The selection of $\alpha$ and $\beta$ was performed against pure random-walk simulations, not numerical experiments (3-body integrations, hence is does not fall under empirical fitting.}
\begin{equation}
\label{eq:beta} 
\beta=2.0.
\end{equation}
\subsection{Comparison of the Random Walk Model to Direct Numerical Integrations}
\label{subsec:comparison}

The results obtained by applying Equations \eqref{eq:NooRW}-\eqref{eq:beta} for the parameters used in  \S\ref{sec:NumericalSimulations} are compared to direct integrations (red dots) in Figure \ref{fig:LifetimeToPericenter}. In both panels $N_{\rm o.o}$ is estimated using either a detail model (Equation \eqref{eq:NooRW}, blue dots) or the rough estimate (Equation \eqref{eq:RoughNooEst}, grey line).  Estimates of $N_{\rm o.o}<1$ are not rounded to 1.  Equation \eqref{eq:TRW} is used to relate $N_{\rm o.o}$ and $T$ for the Random Walk estimates. As can be seen in both panels, the model's predictions show the same overall scatter as the results of 3-body integrations, within about an order of magnitude. An obvious difference is that the Random Walk estimations have tighter and more obvious bounds than the direct integrations. This is expected, as Equations \eqref{eq:NooRW}, \eqref{eq:TRW} represent typical values rather than individual realizations.  

The distribution and medians of the Random Walk model predictions are compared to those of integration in Figure \ref{fig:LifetimeToIncination}, for 5 inclination values and 2 $r_{\rm p.o}$ values. The distribution of predictions made by the model using Equations \eqref{eq:NooRW}-\eqref{eq:alpha} (blue bars) and their medians (blue dashed line) are overlayed on the results of numerical experiments described in \S \ref{sec:NumericalSimulations} (red bars and red dotted lines). The rough estimate of Equation \eqref{eq:RoughNooEst} is plotted in solid grey line. As can be seen, the Random Walk approximations capture the 3-body integrated medians within about an order of magnitude, and shows the same trends of increased $N_{\textrm{o.o}}$ at larger separation and larger inclination values.

An intermediate level of simplification between performing Kozai-Lidov evolution (equation \ref{eq:RWdelta}) and ignoring $e_{\rm i}$, $\iota$ and $\Omega$ altogether (equation \ref{eq:RoughNooEst}) can be obtained by fixing the $e_{\rm o}$ and $\iota$, and averaging over ${\rm M}_{\rm i}^{*}$. This is demonstrated In Figure \ref{fig:NooToIncKozai}, for the same systems used in Figure \ref{fig:LifetimeToIncination}. The red bars, blue bars and gray solid lines are the same as \ref{fig:LifetimeToIncination}, and the Random-Walk model without Kozai-Lidov evolution are in empty bars with dashed exterior. Inclinations $\iota=0,\pi$ are not shown, as the Kozai-Lidov evolution, to quadrupole order, will not change $\iota$ and $e_{\rm i}$ (see appendix \ref{subsec:3BodyBasicsApp}: when $\mathbf{J}_{\rm i}\times\mathbf{J}_{\rm o}=0$ and $\mathbf{e}_{\rm i}\cdot \mathbf{J}_{\rm o}=0$,  $d\mathbf{J}_{\rm i}/dt=0$ and $d\mathbf{e}_{\rm i}/dt \perp \mathbf{e}_{\rm i}$).

\begin{figure}
	\centering
	\includegraphics[width=0.4\textwidth]{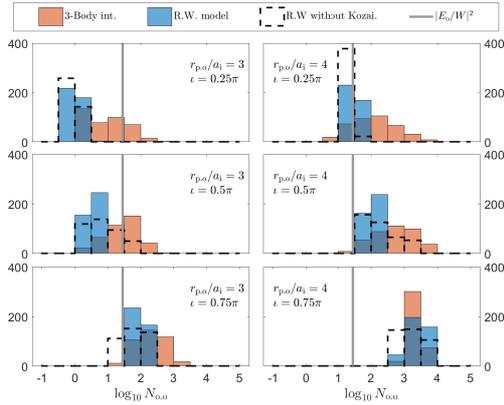}
	\caption{Comparison between the Random Walk model, in three levels of simplification, and direct integration. Systems are same as in Figure \ref{fig:LifetimeToIncination}, without $\iota=0,\pi$. Red bars: 3-body integrations. Blue bars: Random Walk model, for same initial conditions. Empty bars: Random walk model without Kozai Lidov oscillations, uising initial $\iota$, $\Omega$ and $e_{\rm i}$ values. Solid gray line: rough analytic random-walk estimate, ignoring the dependence on orientation and inner eccentricity. }
	\label{fig:NooToIncKozai}
\end{figure}
The comparison shown in Figure \ref{fig:LifetimeToPericenter} is extended to other mass ratios of order unity in Figure \ref{fig:m_1_05_1}  (mass ratios of $m_1:m_2:m_3 = 1:0.5:1$) and Figure \ref{fig:m_1_08_05} ($m_1:m_2:m_3=1:0.8:0.5$). As can be seen, there is little change in the lifetimes compared to the equal mass ratio, in both  direct integrations and Random Walk model. This is expected from Equation \eqref{eq:RoughNooEst}, which implies that $N_{\mathrm{o.o}}$ should change by a factor of a few at most. 
\begin{figure}
	\centering
	\begin{minipage}{0.40\textwidth}
		\includegraphics[width=\textwidth, trim={0 0.2cm 0 1cm}, clip]{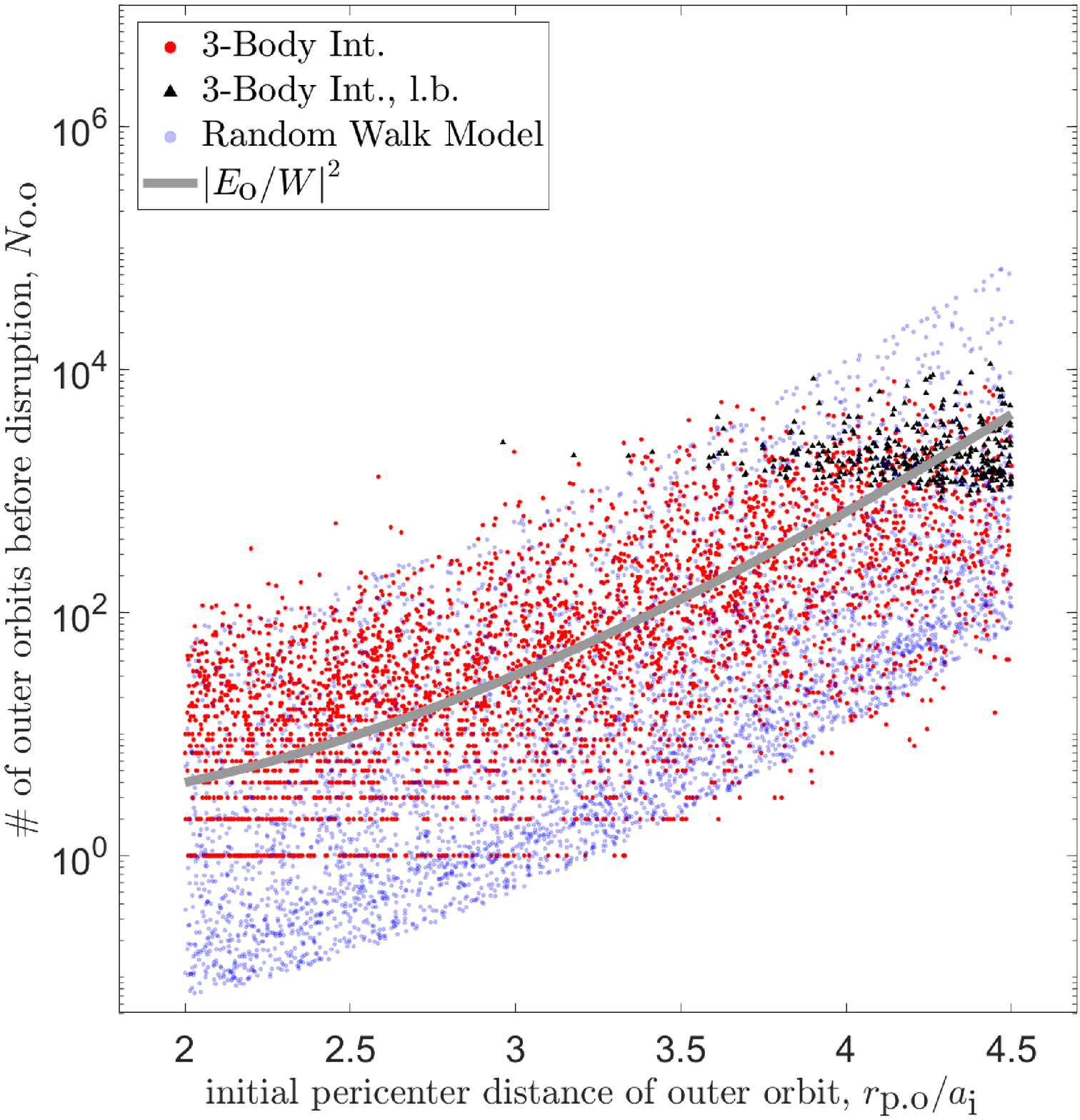}
		\caption{Same as the top panel of Figure \ref{fig:LifetimeToPericenter}, but with $m_2/m_1 = 0.5$, $m_3/m_1=1$. 4,000 simulations performed.}
		\label{fig:m_1_05_1}
	\end{minipage}
	\begin{minipage}{0.40\textwidth}
		\includegraphics[width=\textwidth, trim={0 0.2cm 0 1cm}, clip]{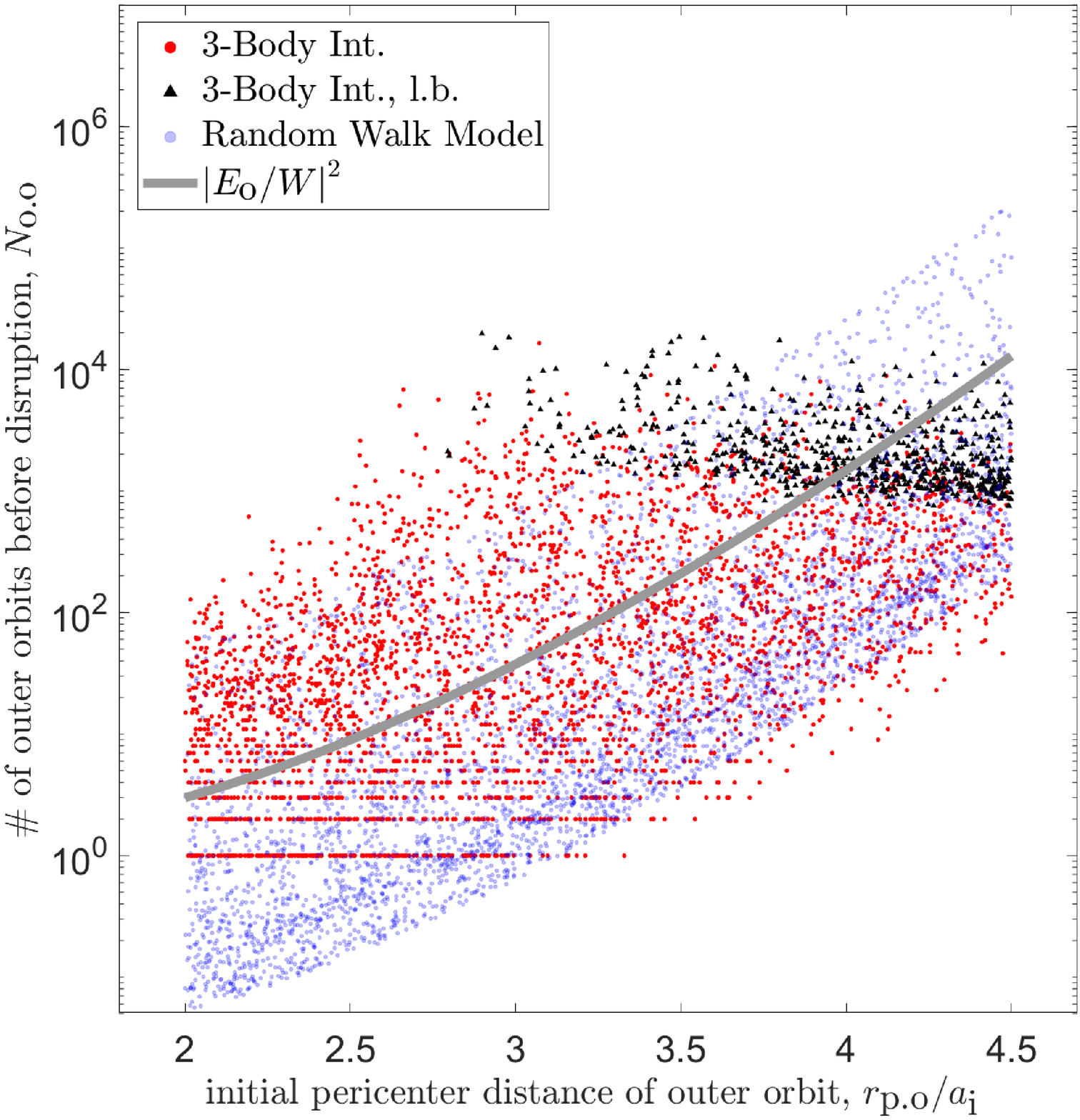}
		\caption{Same as the top panel of Figure \ref{fig:LifetimeToPericenter}, but with $m_2/m_1 = 0.8$, $m_3/m_1=0.5$. 4,000 simulations performed.}
		\label{fig:m_1_08_05}
	\end{minipage}
\end{figure}

While high outer eccentricity is assumed in the Random Walk Model derived in \S \ref{sec:RWmodel}, it is useful to compare it to systems with moderate outer eccentricity, to check the range of its validity. 
Such comparisons to direct numerical integrations are shown in Figures \ref{fig:eo_07}, \ref{fig:eo_03} and \ref{fig:eo_01} ($e_{\rm o}=0.7$, $0.3$ and $0.1$ respectively). In each case, 4,000 3-body systems were integrated, with $r_{\rm p.o}/a_{\rm i}$ drawn randomly from the range $2.0-4.5$. As can be seen, the case of $e_{\rm o}=0.7$ is similar to the case of $e_{\rm o}=0.9$, while the agreement is weaker at lower outer eccentricities. Note that $e_{\mathrm{o}}=0.7$ is very close to the thermal median eccentricity value of $1/\sqrt{2}$, which may be relevant for wide binaries (\citealt{duquennoy_multiplicity_1991}. See however \citealt{raghavan2010survey}). 

\begin{figure}
	\centering
	\begin{minipage}{0.40\textwidth}
		\includegraphics[width=\textwidth, trim={0 0.2cm 0 1cm}, clip]{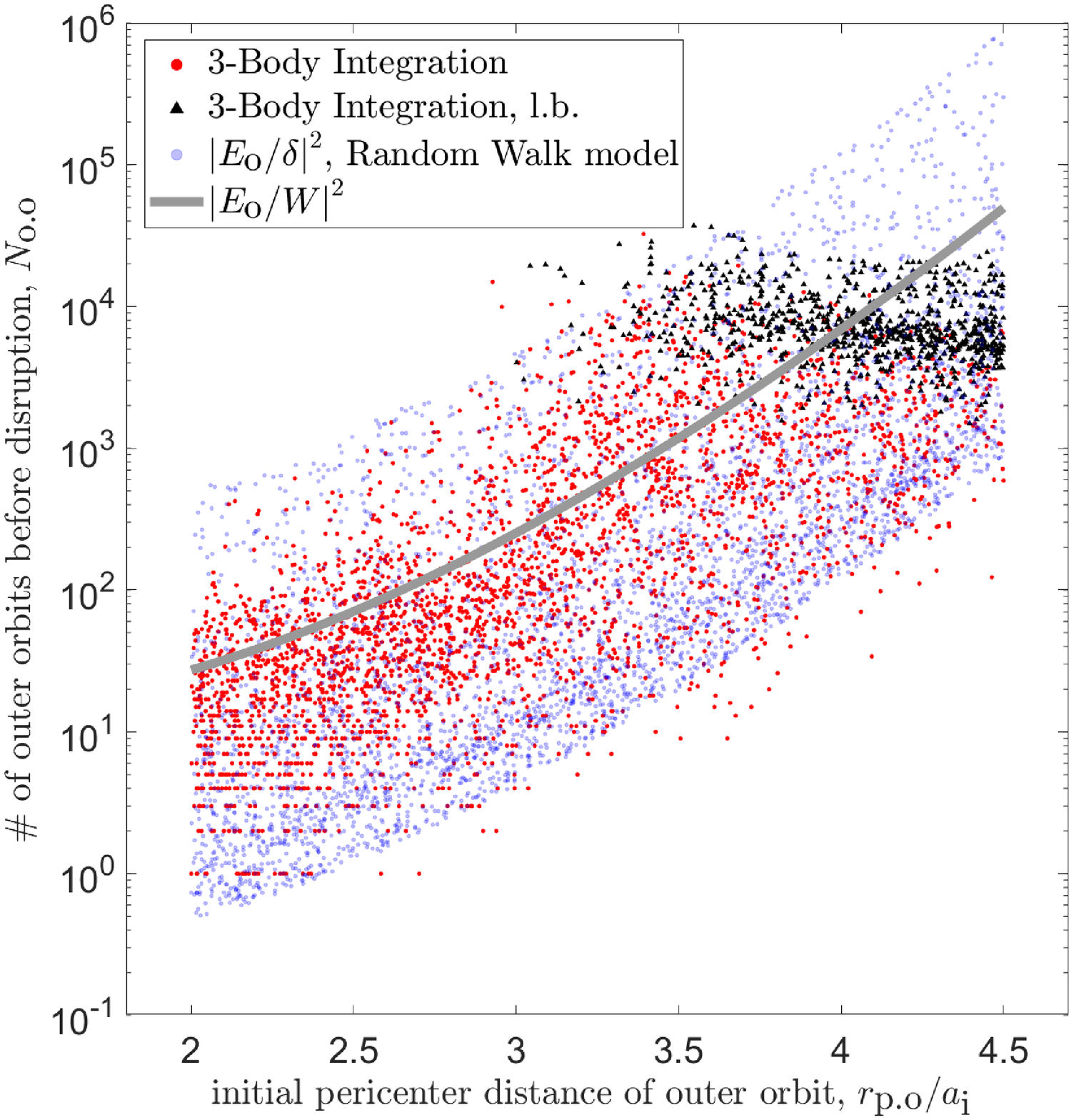}
		\caption{Same as the top panel of Figure \ref{fig:LifetimeToPericenter}, but with $e_{\textrm{o}}=0.7$, close to the thermal mean value of $1/\sqrt{2}$. 4,000 simulations performed.}
		\label{fig:eo_07}
	\end{minipage}
	\begin{minipage}{0.40\textwidth}
		\includegraphics[width=\textwidth, trim={0 0.2cm 0 1cm}, clip]{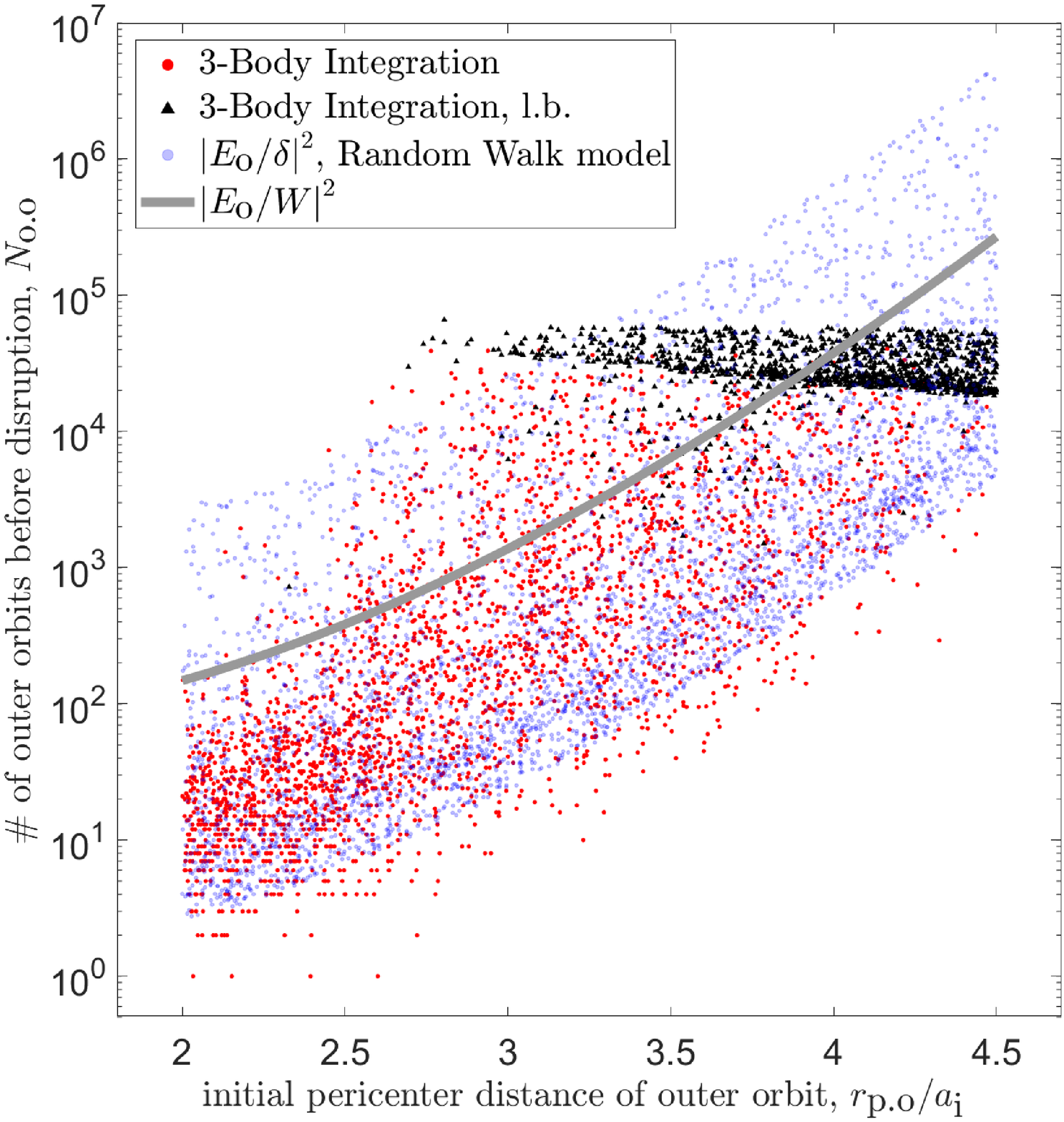}
		\caption{Same as the top panel of Figure \ref{fig:LifetimeToPericenter}, but with $e_{\textrm{o}}=0.3$. 4,000 simulations performed.}
		\label{fig:eo_03}
	\end{minipage}
	\begin{minipage}{0.40\textwidth}	
		\includegraphics[width=\textwidth, trim={0 0.2cm 0 1cm}, clip]{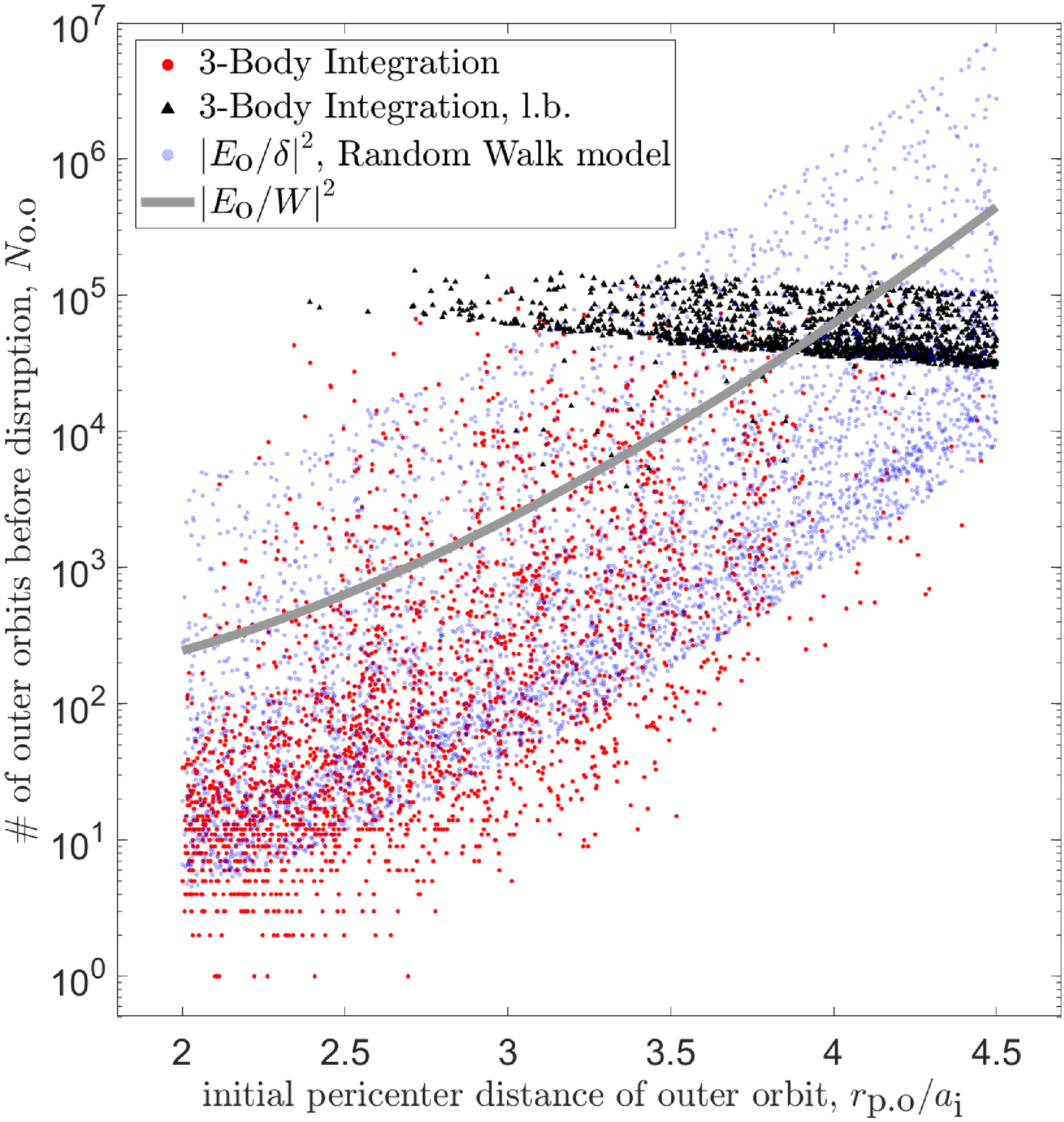}
		\caption{Same as the top panel of Figure \ref{fig:LifetimeToPericenter}, but with $e_{\textrm{o}}=0.1$. 4,000 simulations performed.}
		\label{fig:eo_01}
	\end{minipage}
\end{figure}

%% file: Sections/Conclusions.tex
\section{Summary and Discussion}
\label{sec:summary}
In this paper the disruption process of hierarchical three-body systems with comparable mass and high outer eccentricities was shown to be captured by a simple Random-Walk model in the outer-orbit's energy. In \S \ref{sec:NumericalSimulations} the disruption times of three-body systems were calculated for a wide range of initial conditions using thousands of numerical integrations, employing a dedicated 3-body integration code, see in particular Figures \ref{fig:LifetimeToPericenter}, \ref{fig:LifetimeToIncination} and \ref{fig:m_1_05_1}-\ref{fig:eo_07}. In \S \ref{sec:RWmodel} a simple random-walk model was derived by numerically averaging, over the Kozai-Lidov oscillations, analytic expressions for energy exchanges during outer pericenter passages (approximated as parabolic encounters). The simple model was found to reproduce the numerical disruption times to within an order of magnitude, for a wide range of parameters, as shown in Figures \ref{fig:LifetimeToPericenter} and \ref{fig:LifetimeToIncination}, which are the main results of this paper. As we show, the Kozai-Lidov oscillations do not have a significant effect on the typical disruption time (for a given inclination and random other orientation angles, see Figure \ref{fig:NooToIncKozai}). In fact, a useful rough approximation that ignores the orientation and inner eccentricity can be derived (Equation \ref{eq:RoughNooEst}), and is shown to reproduce the typical disruption times and their dependence on the masses and the hierarchy (see grey lines in Figures \ref{fig:LifetimeToPericenter}, \ref{fig:m_1_05_1}-\ref{fig:eo_07}).

Of the many previous works dealing with the stability of hierarchical three-body systems, a good benchmark would be to the recent work of \cite{myllari2018stability}. As both works are based on a random-walk model and the energy exchange formulas of \cite{roy2003energy}, we expect similar results. Indeed, The dependence of $N_{\rm o.o}$ presented here on $e_{\rm o}$, the masses and $r_{\rm p.o}/a_{\rm i}$ (equation \ref{eq:RoughNooEst}) is the same as in \cite{myllari2018stability}, up to an approximate power-law dependence on $r_{\rm p.o}/a_{\rm i}$ there. Both here and in \citet{myllari2018stability}, larger inclination implies larger $N_{\rm o.o}$ (as shown at Figures \ref{fig:LifetimeToIncination} and \ref{fig:NooToIncKozai}), or phrased as stability criteria, the minimal $r_{\rm p.o}/a_{\rm i}$ for stability is lower for lower $\cos(\iota)$. The Kozai-Lidov mechanism, not incorporated in \cite{myllari2018stability}, was shown not to be significant for most cases, and therefor will rarely alter the stability criterion (as seen by the difference between the lowest $N_{\rm o.o}$ of the Random Walk model with and without Kozai-Lidov evolution, in Figure \ref{fig:NooToIncKozai}).

The duration of the disruption process (see Figure \ref{fig:LifetimeToPericenter}, bottom panel) and the large separation that may be reached, suggests the existence of hierarchical triple systems evolving under prominent galactic tidal disturbance. This scenario, and its observational signature, will be discussed in a follow-up paper. 

Another promising application is to the study of close approaches in multiple systems that can have many astrophysical consequences. Secular Kozai-Lidov calculations performed by \cite{thompson2011accelerating} suggested that the rate of white-dwarf (WD) mergers may be accelerated in hierarchical triple systems, leading to Type Ia supernovae (SNe Ia). \citet{katz2012rate} found that non-secular effects of Kozai-Lidov mechanism in hierarchical triples can produce WD-WD collisions at a possibly comparable rate as SNe Ia. \citet{kushnir2013head} and \citet{dong_type_2015} found evidences supporting WD-WD collisions as a possible major channel for SNe Ia.
Modeling of the disruption process using the results in our work can allow better understanding of the collision probability which is limited by finite
lifetime of the systems \citep[e.g.][]{haim2018extreme}.

%% file: Sections/Appendix1.tex
\section{Derivations}
\subsection{Hierarchical Three Body Systems}
\label{subsec:3BodyBasicsApp}
The motion of bodies in an hierarchical 3-body system can be described through the positions and velocities of the inner and outer orbits, $\bvec{r}_{\textrm{i}},\bvec{r}_{\textrm{o}},\bvec{v}_{\textrm{i}},\bvec{v}_{\textrm{o}}$, that are related to the positions and velocities of the bodies through the Jacobi coordinate transformation. The system's Hamiltonian can be written as a sum of three Hamiltonians, one of the inner binary, one of the outer, and a perturbation term:
\begin{align}
\mathcal{H}=\mathcal{H}_{\textrm{i}} + \mathcal{H}_{\textrm{o}} + \Phi
\end{align}
The perturbation term in the Hamiltonian is respectively given by:
\begin{equation}
\label{eq:InteractionHamiltonian}
\Phi = +\frac{G\mu_{\textrm{o}}M_{\textrm{o}}}{r_{\textrm{o}}}
-
\frac{Gm_{1}m_{3}}{\sqrt{\left(\bvec{r}_{\textrm{o}}+\frac{\mu_{\textrm{i}}}{m_{1}}\bvec{r}_{\textrm{i}}\right)^2}}
-\frac{Gm_{2}m_{3}}{\sqrt{\left(\bvec{r}_{\textrm{o}}-\frac{\mu_{\textrm{i}}}{m_{2}}\bvec{r}_{\textrm{i}}\right)^2}}
\end{equation}
and to second order in $r_{\textrm{i}}/r_{\textrm{o}}$, it it is approximated by:
\begin{equation}
\Phi \approx 
-\frac{Gm_{3}\mu_{\textrm{i}}}{2r_{\textrm{o}}}
\left(1-3\left( \frac{\bvec{r}_{\textrm{i}}}{r_{\textrm{i}}} \cdot \frac{\bvec{r}_{\textrm{o}}}{r_{\textrm{o}}}\right)^2 \right) 
\cdot 
\left(\frac{r_{\textrm{i}}}{r_{\textrm{o}}} \right) ^{2}
+\mathcal{O}\left(\frac{r_{\textrm{i}}}{r_{\textrm{o}}}\right)^{3}
\end{equation}

\subsection{The Double-Averaged Kozai-Lidov Approximation}
\label{subsec:DAKLApp}
Under the Double-Averaging approximation, the 3-body Hamiltonian becomes:
\begin{align}\label{eq:DAHamiltonian}
\left<\mathcal{H}\right> = \mathcal{H}_{\textrm{i}} + \mathcal{H}_{\textrm{o}} + \left<\Phi\right>
\end{align}
with the DA perturbation potential, to quadrupole order, is equal to \citep[Chapter 9]{valtonen2006three}:

\begin{align}\label{eq:KLDA}
\left<\Phi\right> = -\frac{3Gm_{3}\mu_{\textrm{i}}a_{\textrm{i}}^2}{8a_{\textrm{o}}^3(1-e_{\textrm{o}}^2)^{3/2}}
\left( \left(\bvec{j}_{\textrm{i}}\cdot\hat{\bvec{j}}_{\rm o}\right)^2 + 2e_{\textrm{i}}^2 - 5\left(\bvec{e}_{\textrm{i}}\cdot \hat{\bvec{j}}_{\rm o}\right)^2 - \frac{1}{3} \right)
\end{align}
where 
\begin{equation}
\bvec{j} = \frac{\bvec{r} \times \bvec{v} }{\sqrt{GMa}}.
\end{equation}

Given the perturbation Hamiltonian \eqref{eq:KLDA}, one can find how $\left(\bvec{J}_{\textrm{i}}, \bvec{J}_{\textrm{o}}, \bvec{e}_{\textrm{i}}, \bvec{e}_{\textrm{o}}\right)$ evolve with time:
\begin{alignat}{2}
\label{eq:KLdeidt}
\frac{d\bvec{e}_{\textrm{i}}}{dt} &= -\bvec{e}_{\textrm{i}}\times\nabla_{\bvec{J}_{\textrm{i}}}\left<\Phi\right> -\frac{1}{GM_{\textrm{i}}\mu_{\textrm{i}}^2a_{\textrm{i}}} \left(\bvec{J}_{\textrm{i}}\times\nabla_{\bvec{e}_{\textrm{i}}}\left<\Phi\right> \right) \\ 
\label{eq:KLdeodt}
\frac{d\bvec{e}_{\textrm{o}}}{dt} &= -\bvec{e}_{\textrm{o}}\times\nabla_{\bvec{J}_{\textrm{o}}}\left<\Phi\right> -\frac{1}{GM_{\textrm{o}}\mu_{\textrm{o}}^2a_{\textrm{o}}} \left(\bvec{J}_{\textrm{o}}\times\nabla_{\bvec{e}_{\textrm{o}}}\left<\Phi\right> \right) \\
\label{eq:KLdJidt}
\frac{d\bvec{J}_{\textrm{i}}}{dt} &= -\bvec{e}_{\textrm{i}}\times\nabla_{\bvec{e}_{\textrm{i}}}\left<\Phi\right> -\bvec{J}_{\textrm{i}}\times\nabla_{\bvec{J}_{\textrm{i}}}\left<\Phi\right> \\
\label{eq:KLdJodt}
\frac{d\bvec{J}_{\textrm{o}}}{dt} &= -\bvec{e}_{\textrm{o}}\times\nabla_{\bvec{e}_{\textrm{o}}}\left<\Phi\right> -\bvec{J}_{\textrm{o}}\times\nabla_{\bvec{J}_{\textrm{o}}}\left<\Phi\right>
\end{alignat}
Using those equations, we can use numerical integration to calculate the evolution of $\left(\bvec{J}_{\textrm{i}}, \bvec{J}_{\textrm{o}}, \bvec{e}_{\textrm{i}}, \bvec{e}_{\textrm{o}}\right)$.
\subsection{Energy Exchange}
\label{subsec:EnergyExchangeApp}
In a frame of reference set by the plane of motion of the inner binary, $\hat{\bvec{x}}\parallel\bvec{e}_{\mathrm{i}}$ and $\hat{\bvec{z}}\parallel\bvec{J}_\mathrm{i}$ (different from the alignment commonly used in the Kozai-Lidov evolution), the energy change of the outer orbit can be written as Equation (19) in \cite{roy2003energy}, or in Equation (1) in \cite{myllari2018stability}. For simplification, we decompose the expression according to dependences:
\begin{equation}
\label{RHdeltaE}
\begin{aligned}
\Delta E_{\textrm{o}} &= W(a_{\textrm{i}},r_{\textrm{p.o}},m_{1},m_{2},m_{3}) F(\phi,\Omega,\iota,e_{\textrm{i}})
\end{aligned}
\end{equation}
with $W$ being constant during Kozai-Lidov evolution:
\begin{align}
\label{eq:W}
W &= -E_{\textrm{i}} \frac{m_{3}}{M_{\textrm{i}}}\left(\frac{M_{\textrm{i}}}{M_{\textrm{o}}}\right)^{5/4} \left(\frac{r_{\textrm{p.o}}}{a_{\textrm{i}}}\right)^{3/4} e^{-2K/3},\\
K &= \left(\frac{r_{\textrm{p.o}}}{a_{\textrm{i}}}\right)^{3/2}
\left(\frac{2M_{\textrm{i}}}{M_{\textrm{o}}}\right),
\end{align}
$F$ contain the dependencies on orientation (inclination $\iota$ and longitude of ascending node $\Omega$) of the outer orbit relative to the inner one and on the inner eccentricity:
\begin{align}
\label{eq:F}
F = \sqrt{2}A_{1} \sin\phi + 2A_{2}\sin\phi\cos2\Omega + 2 A_{3}\cos\phi\sin2\Omega
\end{align}
with $\phi$ is a defined through the outer argument of periapsis $\omega$ and the mean anomaly of the inner binary at the pericenter approach of the outer orbit $\textrm{M}^{*}_{\textrm{i}}$, calculated for the unperturbed trajectories:
\begin{equation}\label{eq:phi}
\phi = 2\omega- \textrm{M}^{*}_{\textrm{i}}.
\end{equation}
The coefficients $A_{n}(\iota,e_{\textrm{i}})$ are:
\begin{equation}
\label{eq:A1A2A3}
\begin{aligned}
	A_{1} &=\frac{\sqrt{\pi}}{2^{1/4}}(f_2-f_1)\sin^2\iota \\
	A_{2} &=-2^{5/4}\sqrt{\pi}f_4\cos\iota-\frac{\sqrt{\pi}}{2^{7/4}}(f_1+f_2)(3+\cos2\iota)\\
	A_{3} &=-2^{1/4}\sqrt{\pi}(f_1+f_2)\cos\iota-\frac{\sqrt{\pi}}{2^{3/4}}f_4(3+\cos2\iota)
\end{aligned}
\end{equation}
where $f_{1,2,4}$ are functions of the inner eccentricity, which include the Bessel functions:
\begin{equation}
\label{eq:f1f2f4}
\begin{aligned}
f_1 &= J_{-1}\left(e_{\textrm{i}}\right) - 2e_{\textrm{i}} J_{0}\left(e_{\textrm{i}}\right) + 2e_{\textrm{i}}J_{2}\left(e_{\textrm{i}}\right) - J_{3}\left(e_{\textrm{i}}\right) \\
f_2 &= (1-e_{\textrm{i}}^2)\left[ J_{-1}(e_{\textrm{i}}) - J_{3}(e_{\textrm{i}})\right] \\ 
f_4 &= \sqrt{1-e_{\textrm{i}}^2}\left[J_{-1}\left(e_{\textrm{i}}\right) - e_{\textrm{i}} J_{0}\left(e_{\textrm{i}}\right) - e_{\textrm{i}}J_{2}\left(e_{\textrm{i}}\right) + J_{3}\left(e_{\textrm{i}}\right)\right] \\ 
\end{aligned}
\end{equation}

%% file: Sections/Appendix2.tex
\section{Numerical Scheme}
\label{sec:hopon}
\textsc{HopOn} is a PYTHON 3.6 package, tailored for the hierarchical 3-body problem. The package provides tools to create 3-body systems, evolve them numerically and record the systems parameters through time. The core of the code is an Drift-Kick Leapfrog symplectic integrator, from the family of integrators suggested by \cite{preto1999class}. To minimize performance time, \textsc{HopOn} employs the Numba Just-in-Time compiler \citep{lam2015numba}. At sufficiently large separations, the integrator can use the analytical solution to the motion (see section \ref{subsec:KepTimeStep} below). To reduce the output data size, \textsc{HopOn} records the orbital parameters once every outer orbit. The code is available through \href{mailto:jonathan.mushkin@weizmann.ac.il}{jonathan.mushkin@weizmann.ac.il}.

\subsection{3-Body Integration}
\label{subsec:AdaptiveLeapFrog}
In the integrator scheme, \textsc{HopOn} uses adaptive time steps similar to the one used in \citep{katz2012rate}, which is tailored for an hierarchical triplet. Namely, the time-step dependence on potential energy allows it to resolve resolve close pericenter passages. We denote $r,v$ as the physically meaningful position and velocities, and $r^{\textrm{LF}}$ as the leap-frog position, calculated half-way between time-steps.
The equations of motion are:
\begin{alignat}{2}
v_{i+1} &= v_{i} + a(r_{i}^{\textrm{LF}}) \cdot \Delta t_{0} \cdot \left( \frac{ U (r_{i}^{\textrm{LF}}) }{ U_0  }\right)^{-3/2} \label{eq:leap_v2}\\
r_{i+1} &= r_{i}^{\textrm{LF}} + v_{i+1}\cdot\frac{\Delta t_{0}}{2} \cdot \left( \frac{ E_{0} - K(v_{i+1})  }{U_0}  \right)^{-3/2} \label{eq:leap_r_cotemp2}\\
r_{i+1}^{\textrm{LF}} &= r_{i}^{\textrm{LF}} + v_{i+1} \cdot \Delta t_{0} \cdot \left( \frac{ E_{0} - K(v_{i+1})  }{U_0}  \right)^{-3/2} \label{eq:leap_r2}
\end{alignat}
where $a$ is the acceleration, $U_0, E_0$ are the initial potential energy and overall energy of the 3-body system, $K$ is the kinetic energy of the system and $\Delta t_0$ is a constant time selected as a fraction of the initial orbital period of the inner binary:
\begin{equation}
\label{eq:dt0}
\Delta t_0 = \Delta t_{00} \cdot \sqrt{\frac{a_{\textrm{inner}}^3} {G\left(m_{1}+m_{2}\right) } }
\end{equation}
Our control on the integrator resolution is set by selecting $\Delta t_{00}$.  The results presented in this work use a time step coefficient $\Delta t_{00}=0.003$ which is adequate as demonstrated by the convergence test shown in section \ref{sec:convergenceTest} and Figure \ref{fig:ConvergeceTest}. All simulations included a run-time limitation, implemented by stopping the integrations after $10^9$ iterations.


\subsection{Keplerian Timesteps and Termination}
\label{subsec:KepTimeStep}
During a large portion of the outer orbital period, we have small perturbation term $\Phi \sim (r_{\textrm{i}} / r_{\textrm{o}} )^2$  (see Equation \ref{eq:InteractionHamiltonian}). Direct numerical integration will include many time-steps in a portion of the motion which can be accurately approximated as two independent Keplerian motions. To prevent this waste of time, \textsc{HopOn} advances the system analytically as two Keplerian orbits whenever they are sufficiently separated as set by the following criteria. 

We introduce $h\gg 1$ and a lengthscale $L$. Every 100 times steps, the integrator calculates $D_{0}$, the shortest distance between any two bodies, and $D_{1}$, the second shortest distance between any two bodies. For the two bodies involved in $D_0$, it calculates their binary-energy and resulting semi-major axis, $E_{\mathrm{i}}$ and $a_{\mathrm{i}}$. Then, if the following criteria are met, it labels the configuration as hierarchical:
\begin{alignat}{2}
\label{eq:Hcond1} D_0\cdot h &< D_1\\
\label{eq:Hcond2} a_{\mathrm{i}}\cdot h &< D_1\\
\label{eq:Hcond3} L \cdot h &< D_1\\
\label{eq:Hcond4} E_{\mathrm{i}} &<0\\
\label{eq:Hcond5} 0.9 \cdot a_{i} &< D_0
\end{alignat}
Condition \eqref{eq:Hcond1} and \eqref{eq:Hcond2} ensure that the outer orbit is larger in scale than the inner orbit, both instantaneously and throughout the entire inner orbit. Condition \eqref{eq:Hcond3} imposes another fixed lengthscale, to remove pathologies caused by small $a_{\mathrm{i}}$ or numerical deviations in its calculation. Condition \eqref{eq:Hcond4} ensures that inner orbit is bound, and condition \eqref{eq:Hcond5} makes sure that the inner binary is instantaneously separated enough so that energy calculation errors are not significant. Those precautions are taken as the adaptive-timestep leapfrog integrator does not strictly conserve energy.

If the configuration is hierarchical, the integrator will decide if to perform a Keplerian time-step, terminate the simulation, or perform a regular leapfrog time-step. To answer this question, it calculates the outer energy. If
\begin{equation}
\label{eq:KepCond} E_{\mathrm{o}}<0\quad\mathrm{and}\quad \bvec{r}_{\mathrm{o}}\cdot\bvec{v}_{\mathrm{o}}>0
\end{equation}
then a Keplerian time-step is performed. If 
\begin{equation}
\label{eq:TerminationCond} E_{\mathrm{o}}>0\quad\mathrm{and}\quad \bvec{r}_{\mathrm{o}}\cdot\bvec{v}_{\mathrm{o}}>0
\end{equation}
then the simulation can be terminated, as the third body is bound to go to infinity. If the system is not hierarchical, or if neither conditions \eqref{eq:KepCond} nor \eqref{eq:TerminationCond} are met, it performs a leapfrog step. 
Explicitly, we choose in our simulations to use $h=100$, and $L=a_{\textrm{i}}^{(t=0)}$. 
\subsection{Convergence Test}
\label{sec:convergenceTest}
Convergence of an individual simulation is often practically impossible to achieve once the integration time is longer than the Lyapunov time of the system \citep[Chapter 2]{valtonen2006three}. Even the slightest numerical disagreement can propagate into the significant digits within the integration run time. Convergence can only be tested in the statistical sense, in claims made about many simulations. We perform this by repeating the calculation of median $N_{\rm o.o}$ (as presented in red dotted lines in Figure \ref{fig:LifetimeToIncination}) with varying numerical resolution.

\begin{figure}
	\centering
	\includegraphics[width=0.5\textwidth]{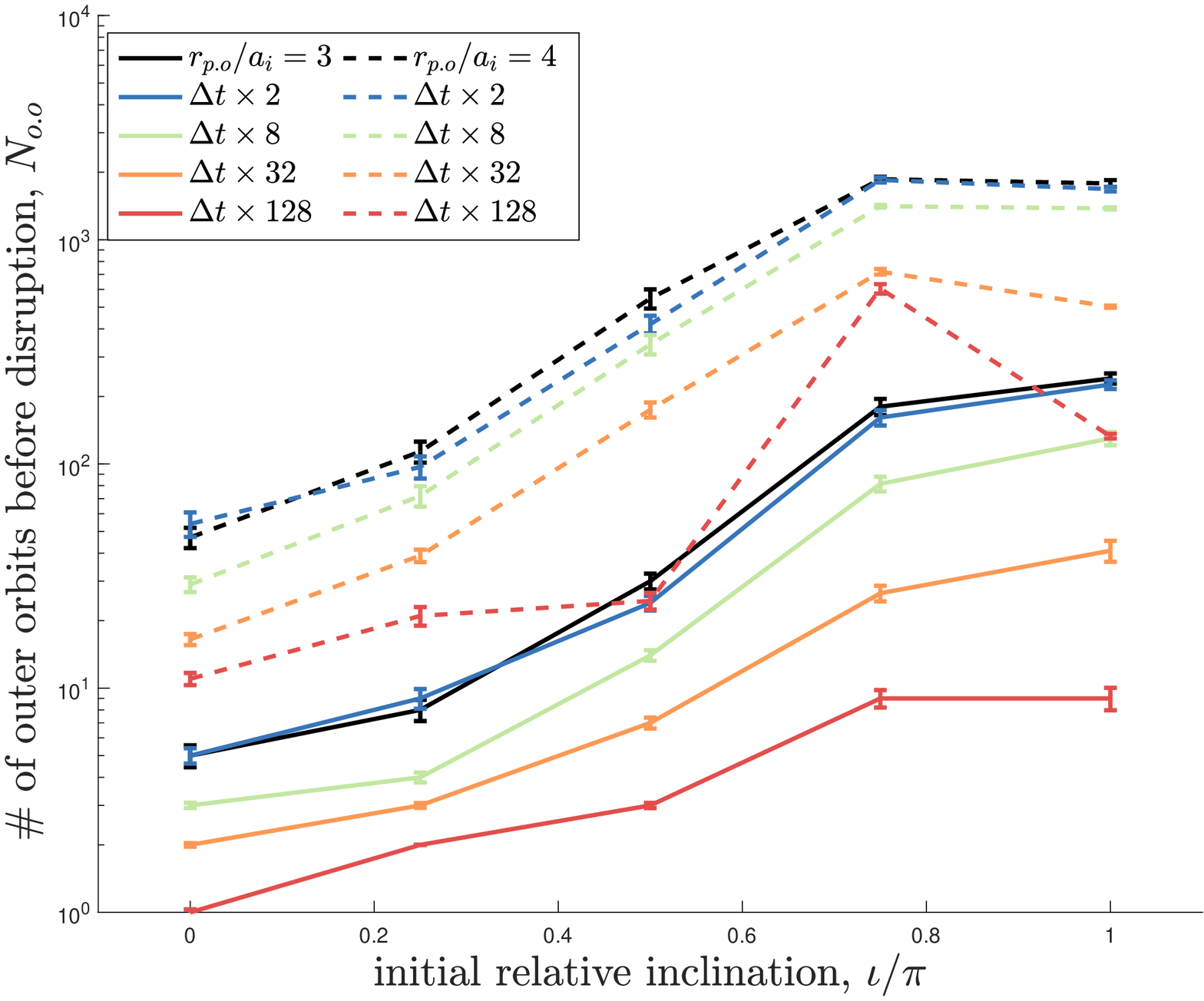}
	\caption{Median number of outer orbits completed before disruption, plotted against inclination, for two fixed $r_{\mathrm{p.o}}$ values. Each point on curves represent the an ensemble of 400 simulations of hierarchical triplets with random orientation and fixed $r_{\textrm{p.o}}$ and $\iota$, specified in table \ref{tab:ic}.  Black solid line: simulations with $\Delta t_{00}=0.003$, and $r_{\textrm{p.o}}/a_{\textrm{i}}=3$ . Colored solid lines: simulations with same initial conditions as in the black line, but with basic time step $\Delta t_{00}$ (Equation \ref{eq:dt0}) larger by factor of 2 (blue), 8 (green), 32 (orange) or 128 (red), and iteration constraint smaller by same factors. Dashed lines: same as solid lines, but with $r_{\textrm{p.o}}/a_{\textrm{i}}=4$.  } 
	\label{fig:ConvergeceTest}
\end{figure}

We performed 4,000 numerical experiments of hierarchical triplets, with eccentric outer orbit ($e_{\textrm{o}}=0.9$), slightly eccentric inner binary ($e_{\textrm{i}}=0.5$), equal masses ($m_1=m_2=m_3$). They are divided into 10 batches of 400 simulations of equal inclination and $r_{\textrm{p.o}}$, chosen from the 10 possible pairs pairs ($r_{\textrm{p.o}}/a_{\textrm{i}}=$3, 4, $\iota=0$, $\pi/4$, $\pi/2$, $3\pi/4$ and $\pi$). In each simulation, the inner and outer orbits has random relative phase, and isotropic random relative angles $\Omega$ and $\omega$. For each $\iota$-$r_{\textrm{p.o}}$ pair, we found the median number of completed outer orbits before disruption. This was repeated, increasing $\Delta t_{00}$ (defined in Equation \ref{eq:dt0}) from its initial value of 0.003 by factor of 2, 8, 32, or 128, and reducing the run-time constraint of $10^9$ iterations by same factors. In Figure \ref{fig:ConvergeceTest} we present the convergence curves. Error bars represent usual median estimation asymptotic standard deviation, with probability density at the median approximated using the 40 and 60 percentiles of the sample. Convergence is visible in the sense that the curves become more crowded as the multiplicative factor on $\Delta t_{00}$ is reduces. 

%% file: Sections/Appendix3.tex
\section{Initial Conditions for Simulations}
\label{sec:ICApp}
Specification of the initial conditions of all simulations performed for this work. Angles $\iota$, $\Omega$ and $\omega$ (without subscripts) are of the outer orbit, in a coordinate system set by the inner orbit, with $\hat{\bvec{x}}\parallel \bvec{e}_{\rm i}$ and $\hat{\bvec{z}}\parallel\bvec{J}_{\rm i}$.

\begin{table*}
\centering
\caption{Initial conditions for simulations used in this work. }
\label{tab:ic}
\begin{tabular}{|l| l| l| l| l| l| l| l| l| l| l| l| l| l| }
	\hline
	Experiment & $N_{\mathrm{simulations}}$ &
	$m_1$ & $m_2$ & $m_3$ & 
	$a_{\mathrm{i}}$ & $r_{\mathrm{p.o}}$ & $e_{\mathrm{i}}$ &	 
	$e_{\mathrm{o}}$ & $\mathrm{M}_{\mathrm{o}}$ & $\mathrm{M}_{\mathrm{i}}$ & 
	$\Omega$ & $\omega$	&  $\cos(\iota)$ \\
	\hline
	
	Figure \ref{fig:LifetimeToPericenter} & 4000 & 
	1 & 1 & 1 & 
	1 & $\mathcal{U}(2.0,4.5)$ & 0.5 & 
	0.9 & $\pi$ & $\mathcal{U}(0,2\pi)$ & 
	$\mathcal{U}(0,2\pi)$ & $\mathcal{U}(0,2\pi)$ & $\mathcal{U}(-1,1)$\\
	
	Figures \ref{fig:LifetimeToIncination}, \ref{fig:NooToIncKozai}, \ref{fig:ConvergeceTest} & 2$\times$5$\times$400 &
	1 & 1 & 1 & 
	1 & 3, 4 & 0.5 & 
	0.9 & $\pi$ & $\mathcal{U}(0,2\pi)$ & 
	$\mathcal{U}(0,2\pi)$ & $\mathcal{U}(0,2\pi)$ & $\frac{\pi}{4}k$, $k=0,1,2,3,4$\\
	
	Figure \ref{fig:3BodyExample} & 1 &
	1 & 1 & 1 & 
	1 & 3.74 & 0.5 & 
	0.9 & $\pi$ & 4.0582 & 
	5.2848 & 5.4148 & $\cos(1.44)$\\

	Figure \ref{fig:DeltaEoVarianceDemonstration} & 10000 & 
	1 & 1 & 1 & 
	1 & 4.0 & 0.5 & 
	0.9 & $\pi$ & $\mathcal{U}(0,2\pi)$ & 
	$0.2\pi$ & $1.5\pi$ & $-\frac{1}{\sqrt{2}}$\\
	
		Figure \ref{fig:m_1_05_1} & 4000 & 
	1 & 0.5 & 1 & 
	1 & $\mathcal{U}(2.0,4.5)$ & 0.5 & 
	0.9 & $\pi$ & $\mathcal{U}(0,2\pi)$ & 
	$\mathcal{U}(0,2\pi)$ & $\mathcal{U}(0,2\pi)$ & $\mathcal{U}(-1,1)$\\
	
	Figure \ref{fig:m_1_08_05} & 4000 & 
	1 & 0.8 & 0.5 & 
	1 & $\mathcal{U}(2.0,4.5)$ & 0.5 & 
	0.9 & $\pi$ & $\mathcal{U}(0,2\pi)$ & 
	$\mathcal{U}(0,2\pi)$ & $\mathcal{U}(0,2\pi)$ & $\mathcal{U}(-1,1)$\\
	
	Figure \ref{fig:eo_07} & 4000 & 
	1 & 1 & 1 & 
	1 & $\mathcal{U}(2.0,4.5)$ & 0.5 & 
	0.7 & $\pi$ & $\mathcal{U}(0,2\pi)$ & 
	$\mathcal{U}(0,2\pi)$ & $\mathcal{U}(0,2\pi)$ & $\mathcal{U}(-1,1)$\\
	
	Figure \ref{fig:eo_03} & 4000 & 
	1 & 1 & 1 & 
	1 & $\mathcal{U}(2.0,4.5)$ & 0.5 & 
	0.3 & $\pi$ & $\mathcal{U}(0,2\pi)$ & 
	$\mathcal{U}(0,2\pi)$ & $\mathcal{U}(0,2\pi)$ & $\mathcal{U}(-1,1)$\\
	
	Figure \ref{fig:eo_01} & 4000 & 
	1 & 1 & 1 & 
	1 & $\mathcal{U}(2.0,4.5)$ & 0.5 & 
	0.1 & $\pi$ & $\mathcal{U}(0,2\pi)$ & 
	$\mathcal{U}(0,2\pi)$ & $\mathcal{U}(0,2\pi)$ & $\mathcal{U}(-1,1)$\\

\hline
\end{tabular}
\end{table*}